\documentstyle[12pt]{article}
\title{Current tensor with heavy photon for hard pair production
by longitudinally polarized electron}

\author{M.Konchatnij, N.P.Merenkov, O.N.Shekhovzova}

\date{}
\begin{document}
\maketitle
\begin{center}
{\it Institute of Physics and Technology, Kharkov, 310108,
Ukraine}
%$^{2)}$ \it Dipartamento di Fisica, Universit\'a di Parma and INFN,  \\
%Gruppo Collegato di Parma, 43100, Parma, Italy}
\\[.2cm]
\end{center}

\vspace{0.5cm}
\begin{abstract}
The electron current tensor for the scattering of heavy photon on
longitudinally polarized electron accompanied with additional hard
electron--positron pair has been considered. The contribution of
collinear and semicollinear kinematics is computed. The full analysis of
both, spin--independent and spin--dependent parts of electron current
tensor, is performed. The obtained result allows to calculate the
corresponding contribution into the second order radiative correction to
cross--sections of different processes with the next--to--leading
accuracy.  \end{abstract}
\section{Introduction}

\hspace{0.6cm}The recent polarized experiments on deep inelastic
scattering \cite{SMC,HERMES} cover the kinematical region
of the Bjorken variable $y\simeq0.9$,
where the electromagnetic corrections to the cross--section are extremely
large. The corresponding first--order QED correction due to single real
and virtual photon emission have been computed in \cite{ksh,ash}, and this
correction at large values of variable $y$ is of the order the Born
cross--section. Therefore, the calculation of the
second--order QED correction becames very important for interpretation of
these
experiments in terms of hadron structure functions.

The DIS cross--section in general case can be represented as a product
of the electron current tensor (ECT) and the hadron one \cite{RF}.
The ECT is model--independent and universal, while the hadron one
depends on used model for description of hadron and has its own
specification for different processes, event selection and so on. That is
why it is very important to calculate the universal quantity ECT with
maximal possible accuracy, because it can be applied to many processes in
scattering and annihilation channels.

The first steps in calculation of the second--order QED correction to
the ECT were done in \cite{AAK98,KM98}. In \cite{AAK98} the
one--loop corrected Compton tensor with a heavy photon was calculated in
the limited case $m=0$, where $m$ is the electron mass. Such kind of
approximation does not take into account all contribution with
next--to--leading accuracy if radiated photon is not observed. In \cite{KM98} the
ECT due to hard double--photon emission have been derived
keeping the electron mass finite and taking into account the contributions
of collinear and semicollinear kinematics.  Just such kind of
approximation allows to keep all next--to--leading terms for the case
of unobserved photons.

Except an additional contribution into ECT due to virtual and
real soft double--photon emission and $e^+e^-$--pair production, there is
the contribution due to hard pair production, and in this work we
calculate it with the same accuracy as it was done in \cite{KM98} for hard
double--photon emission.

In the Born approximation the ECT  with longitudinally
polarized electron has the form
\begin{equation}\label{1Born}
L_{\mu\nu}^{^B} = Q_{\mu\nu} + i\lambda E_{\mu\nu}\ , \ \ Q_{\mu\nu} =
-4(p_1p_2)g_{\mu\nu} + 4p_{1\mu}p_{2\nu} +4p_{1\nu}p_{2\mu}\ ,
\end{equation}
$$E_{\mu\nu} = 4\epsilon_{\mu\nu\rho\sigma}p_{1\rho}p_{2\sigma} \ ,$$
where $p_1(p_2)$ is the 4--momentum of the initial (final) electron, and
$\lambda $ is the doubled initial electron helicity. (The value
$\lambda$ equals 1 (or -1) if the initial electron is polarized
along (against) its 3--momentum direction).

Here we consider the corrections to the tensor $L_{\mu\nu}^{^B}$
due to pair production in the scattered
\begin{equation}\label{2,process}
e^-(p_1) + \gamma^{^*}(q) \rightarrow e^-(p_2) + e^+(p_+) + e^-(p_-)
\end{equation}
or annihilation
$$e^-(p_1) + e^+(p_2) \rightarrow  \gamma^{^*}(q) + e^-(p_-) + e^+(p_+)$$
processes under the condition
$$|q^2|\ , \ (p_1p_2) \gg m^2 $$
in collinear and semicollinear kinematics. The corresponding calculations
for unpolarized case were performed in part for DIS
[8,9] as well as for small-- and large--angle Bhabha
scattering [10,11] processes on the level of cross--sections. Some
other aspects of QED corrections due to pair production are discussed
also in \cite{BNB88,KPF91Nov,ESF,AKMTSP97}. For definiton, we
will investigate the scattering channel, and, to obtain the corresponding
result for the annihilation one, it needs to substitute $-p_2$
instead of $p_2.$

The paper is organized as follows. In Section 2 we consider the
contribution due to collinear kinemtics. There are two collinear
kinematical regions:  $a)$ -- when created pair is emitted along the
initial electron momentum direction $(\vec p_+\ , \vec p_-\ \parallel\vec
p_1)$, and $b)$ -- when created pair flies along the final one $(\vec p_+\
, \vec p_-\ \parallel\vec p_2).$ In collinear regions both, photon and
fermion proparator denominators (PD)
of underlying essential Feynman diagrams, can be
small. The corresponding contribution into ECT can be expressed in terms
of symmetrical tensor $Q_{\mu\nu}$ and antisymmetrical tensor $E_{\mu\nu}$
in the same manner as it was done in \cite{KM98} for double--photon
emission process.  In Section 2.1 we calculate the ECT in the region $a)$,
and in Section 2.2 -- in the region $b).$

In Section 3 we investigate the ECT in semicollinear kinematics. Only
photon PD's
of Feynman diagrams can be small in this case, and under the
considered here conditions there are three
different semicollinear regions \cite{NPM89}: $(\vec p_-\ \parallel\vec p_1)\ ,
(\vec p_-\ \parallel\vec p_+)$, and $(\vec p_+\ \parallel\vec p_2).$ The
corresponding contributions into ECT we calculate in Sections 3.1, 3.2
and 3.3, respectively. The structure of the ECT in semicollinear regions
is more complicated as compared with collinear ones. We demonstrate the
elimination of the angular auxiliary parameters used to define
collinear regions $a)$ and $b)$ in sum of contributions due to collinear
and semicollinear kinematics on the level of next--to--leading accuracy.
The starting--points of our calculations given in
Appendix.  The main obtained results are formulated briefly in Conclusion.

\section{Investigation of collinear kinematics}

\hspace{0.6cm}As we noted in Introduction in collinear region both, photon
and electron propagator denominators of underlying Feynman diagrams, can
be small.  Because of used restriction on the mass of the heavy photon and
registration condition for the scattered electron only six from the whole
eight diagrams set are essential. These diagrams are shown on Fig.1.

%-----------------------fig.1---------------------------------
\begin{figure}[t]
\unitlength=0.70mm
\special{em:linewidth 0.6pt}
\linethickness{0.6pt}
\begin{picture}(190.00,65.00)
%-----------------------(1,2)-----------------------------
\put(05.00,30.00){\vector(1,0){10.00}}
\put(15.00,30.00){\line(1,0){10.50}}
\put(08.00,33.00){\makebox(0,0)[cc]{$p_1$}}
\put(30.00,30.00){\oval(9.00,9.00)[]}
\put(34.50,30.00){\vector(1,0){10.50}}
\put(45.00,30.00){\line(1,0){10.00}}
\put(52.00,33.00){\makebox(0,0)[cc]{$p_2$}}
\put(30.00,24.00){\oval(3.00,3.00)[r]}
\put(30.00,21.00){\oval(3.00,3.00)[l]}
\put(30.00,18.00){\oval(3.00,3.00)[r]}
\put(30.00,15.00){\oval(3.00,3.00)[l]}
\put(30.00,12.00){\oval(3.00,3.00)[r]}
\put(30.00,10.50){\makebox(0,0)[cc]{$\times$}}
\put(30.00,02.00){\makebox(0,0)[cc]{(1,2)}}
\put(32.00,34.50){\oval(4.00,4.00)[lt]}
\put(32.00,38.50){\oval(4.00,4.00)[rb]}
\put(36.00,38.50){\oval(4.00,4.00)[lt]}
\put(36.00,42.50){\oval(4.00,4.00)[rb]}
\put(38.00,42.50){\line(1,4){2.00}}
\put(42.00,58.50){\vector(-1,-4){2.00}}
\put(35.00,56.00){\makebox(0,0)[cc]{$-p_{+}$}}
\put(38.00,42.50){\vector(4,1){8.00}}
\put(46.00,44.50){\line(4,1){8.00}}
\put(50.50,48.00){\makebox(0,0)[cc]{$p_{-}$}}
\put(65.00,30.00){\makebox(0,0)[cc]{$-$}}
%-----------------------(3,4)-----------------------------
\put(75.00,30.00){\vector(1,0){10.00}}
\put(85.00,30.00){\line(1,0){10.50}}
\put(78.00,33.00){\makebox(0,0)[cc]{$p_1$}}
\put(100.00,30.00){\oval(9.00,9.00)[]}
\put(104.50,30.00){\vector(1,0){10.50}}
\put(115.00,30.00){\line(1,0){10.00}}
\put(122.00,33.00){\makebox(0,0)[cc]{$p_{-}$}}
\put(100.00,24.00){\oval(3.00,3.00)[r]}
\put(100.00,21.00){\oval(3.00,3.00)[l]}
\put(100.00,18.00){\oval(3.00,3.00)[r]}
\put(100.00,15.00){\oval(3.00,3.00)[l]}
\put(100.00,12.00){\oval(3.00,3.00)[r]}
\put(100.00,10.50){\makebox(0,0)[cc]{$\times$}}
\put(100.00,02.00){\makebox(0,0)[cc]{(3,4)}}
\put(102.00,34.50){\oval(4.00,4.00)[lt]}
\put(102.00,38.50){\oval(4.00,4.00)[rb]}
\put(106.00,38.50){\oval(4.00,4.00)[lt]}
\put(106.00,42.50){\oval(4.00,4.00)[rb]}
\put(108.00,42.50){\line(1,4){2.00}}
\put(112.00,58.50){\vector(-1,-4){2.00}}
\put(105.00,56.00){\makebox(0,0)[cc]{$-p_{+}$}}
\put(108.00,42.50){\vector(4,1){8.00}}
\put(116.00,44.50){\line(4,1){8.00}}
\put(120.50,48.00){\makebox(0,0)[cc]{$p_{2}$}}
\put(135.00,30.00){\makebox(0,0)[cc]{$-$}}
%-----------------------(5,6)-----------------------------
\put(165.00,30.00){\oval(9.00,9.00)[]}
\put(165.00,24.00){\oval(3.00,3.00)[r]}
\put(165.00,21.00){\oval(3.00,3.00)[l]}
\put(165.00,18.00){\oval(3.00,3.00)[r]}
\put(165.00,15.00){\oval(3.00,3.00)[l]}
\put(165.00,12.00){\oval(3.00,3.00)[r]}
\put(165.00,10.50){\makebox(0,0)[cc]{$\times$}}
\put(165.00,36.00){\oval(3.00,3.00)[r]}
\put(165.00,39.00){\oval(3.00,3.00)[l]}
\put(165.00,42.00){\oval(3.00,3.00)[r]}
\put(165.00,45.00){\oval(3.00,3.00)[l]}
\put(165.00,48.00){\oval(3.00,3.00)[r]}
\put(145.00,49.50){\vector(1,0){10.00}}
\put(155.00,49.50){\vector(1,0){20.00}}
\put(175.00,49.50){\line(1,0){10.00}}
\put(148.00,52.50){\makebox(0,0)[cc]{$p_1$}}
\put(182.00,52.50){\makebox(0,0)[cc]{$p_{-}$}}
\put(169.00,33.00){\vector(1,0){8.00}}
\put(177.00,33.00){\line(1,0){08.00}}
\put(182.00,36.00){\makebox(0,0)[cc]{$p_{2}$}}
\put(185.00,27.00){\vector(-1,0){10.00}}
\put(169.00,27.00){\line(1,0){08.00}}
\put(180.00,24.00){\makebox(0,0)[cc]{$-p_{+}$}}
\put(165.00,02.00){\makebox(0,0)[cc]{(5,6)}}
\end{picture}

\begin{picture}(190.00,50.00)
\put(05.00,30.00){\vector(1,0){10.00}}
\put(15.00,30.00){\line(1,0){10.50}}
\put(30.00,30.00){\oval(9.00,9.00)[]}
\put(34.50,30.00){\vector(1,0){10.50}}
\put(45.00,30.00){\line(1,0){10.00}}
\put(30.00,24.00){\oval(3.00,3.00)[r]}
\put(30.00,21.00){\oval(3.00,3.00)[l]}
\put(30.00,18.00){\oval(3.00,3.00)[r]}
\put(30.00,15.00){\oval(3.00,3.00)[l]}
\put(30.00,12.00){\oval(3.00,3.00)[r]}
\put(30.00,10.50){\makebox(0,0)[cc]{$\times$}}
\put(32.00,34.50){\oval(4.00,4.00)[lt]}
\put(32.00,38.50){\oval(4.00,4.00)[rb]}
\put(36.00,38.50){\oval(4.00,4.00)[lt]}
\put(36.00,42.50){\oval(4.00,4.00)[rb]}
\put(65.00,30.00){\makebox(0,0)[cc]{$=$}}
%----------------------------------------------------
\put(75.00,30.00){\vector(1,0){8.00}}
\put(83.00,30.00){\vector(1,0){18.00}}
\put(99.00,30.00){\line(1,0){8.00}}
\put(87.00,30.00){\oval(4.00,4.00)[lt]}
\put(87.00,34.00){\oval(4.00,4.00)[rb]}
\put(91.00,34.00){\oval(4.00,4.00)[lt]}
\put(91.00,38.00){\oval(4.00,4.00)[rb]}
\put(95.00,38.00){\oval(4.00,4.00)[lt]}
\put(95.00,42.00){\oval(4.00,4.00)[rb]}
\put(93.00,25.50){\oval(3.00,3.00)[l]}
\put(93.00,28.50){\oval(3.00,3.00)[r]}
\put(93.00,25.50){\oval(3.00,3.00)[l]}
\put(93.00,22.50){\oval(3.00,3.00)[r]}
\put(93.00,19.50){\oval(3.00,3.00)[l]}
\put(93.00,16.50){\oval(3.00,3.00)[r]}
\put(93.00,13.50){\oval(3.00,3.00)[l]}
\put(93.00,12.00){\makebox(0,0)[cc]{$\times$}}
\put(117.00,30.00){\makebox(0,0)[cc]{$+$}}
%----------------------------------------------------
\put(127.00,30.00){\vector(1,0){8.00}}
\put(135.00,30.00){\vector(1,0){18.00}}
\put(151.00,30.00){\line(1,0){8.00}}
\put(145.00,30.00){\oval(4.00,4.00)[lt]}
\put(145.00,34.00){\oval(4.00,4.00)[rb]}
\put(149.00,34.00){\oval(4.00,4.00)[lt]}
\put(149.00,38.00){\oval(4.00,4.00)[rb]}
\put(153.00,38.00){\oval(4.00,4.00)[lt]}
\put(153.00,42.00){\oval(4.00,4.00)[rb]}
\put(137.00,25.50){\oval(3.00,3.00)[l]}
\put(137.00,28.50){\oval(3.00,3.00)[r]}
\put(137.00,25.50){\oval(3.00,3.00)[l]}
\put(137.00,22.50){\oval(3.00,3.00)[r]}
\put(137.00,19.50){\oval(3.00,3.00)[l]}
\put(137.00,16.50){\oval(3.00,3.00)[r]}
\put(137.00,13.50){\oval(3.00,3.00)[l]}
\put(137.00,12.00){\makebox(0,0)[cc]{$\times$}}
\end{picture}

\begin{picture}(190.00,50.00)
\put(10.00,30.00){\oval(9.00,9.00)[]}
\put(14.00,33.00){\vector(1,0){8.00}}
\put(22.00,33.00){\line(1,0){08.00}}
\put(30.00,27.00){\vector(-1,0){10.00}}
\put(14.00,27.00){\line(1,0){08.00}}
\put(10.00,24.00){\oval(3.00,3.00)[r]}
\put(10.00,21.00){\oval(3.00,3.00)[l]}
\put(10.00,18.00){\oval(3.00,3.00)[r]}
\put(10.00,15.00){\oval(3.00,3.00)[l]}
\put(10.00,12.00){\oval(3.00,3.00)[r]}
\put(10.00,10.50){\makebox(0,0)[cc]{$\times$}}
\put(10.00,36.00){\oval(3.00,3.00)[r]}
\put(10.00,39.00){\oval(3.00,3.00)[l]}
\put(10.00,42.00){\oval(3.00,3.00)[r]}
\put(10.00,45.00){\oval(3.00,3.00)[l]}
\put(10.00,48.00){\oval(3.00,3.00)[r]}
\put(10.00,49.50){\makebox(0,0)[cc]
%{$\times$}
}
\put(40.00,30.00){\makebox(0,0)[cc]{$=$}}
%----------------------------------------------------
\put(50.00,34.50){\line(0,-1){09.00}}
\put(86.00,25.50){\vector(0,1){06.00}}
\put(50.00,34.50){\vector(1,0){8.00}}
\put(58.00,34.50){\line(1,0){08.00}}
\put(66.00,25.50){\vector(-1,0){10.00}}
\put(50.00,25.50){\line(1,0){08.00}}
\put(50.00,24.00){\oval(3.00,3.00)[r]}
\put(50.00,21.00){\oval(3.00,3.00)[l]}
\put(50.00,18.00){\oval(3.00,3.00)[r]}
\put(50.00,15.00){\oval(3.00,3.00)[l]}
\put(50.00,12.00){\oval(3.00,3.00)[r]}
\put(50.00,10.50){\makebox(0,0)[cc]{$\times$}}
\put(50.00,36.00){\oval(3.00,3.00)[r]}
\put(50.00,39.00){\oval(3.00,3.00)[l]}
\put(50.00,42.00){\oval(3.00,3.00)[r]}
\put(50.00,45.00){\oval(3.00,3.00)[l]}
\put(50.00,48.00){\oval(3.00,3.00)[r]}
\put(50.00,49.50){\makebox(0,0)[cc]
%{$\times$}
}
\put(76.00,30.00){\makebox(0,0)[cc]{$+$}}
%----------------------------------------------------
\put(86.00,34.50){\line(0,-1){09.00}}
\put(50.00,34.50){\vector(0,-1){06.00}}
\put(86.00,25.50){\vector(2,1){16.00}}
\put(86.00,25.50){\line(2,1){18.00}}
\put(104.00,25.50){\vector(-2,1){04.00}}
\put(86.00,34.50){\line(2,-1){18.00}}
\put(86.00,24.00){\oval(3.00,3.00)[r]}
\put(86.00,21.00){\oval(3.00,3.00)[l]}
\put(86.00,18.00){\oval(3.00,3.00)[r]}
\put(86.00,15.00){\oval(3.00,3.00)[l]}
\put(86.00,12.00){\oval(3.00,3.00)[r]}
\put(86.00,10.50){\makebox(0,0)[cc]{$\times$}}
\put(86.00,36.00){\oval(3.00,3.00)[r]}
\put(86.00,39.00){\oval(3.00,3.00)[l]}
\put(86.00,42.00){\oval(3.00,3.00)[r]}
\put(86.00,45.00){\oval(3.00,3.00)[l]}
\put(86.00,48.00){\oval(3.00,3.00)[r]}
\put(86.00,49.50){\makebox(0,0)[cc]
%{$\times$}
}
\end{picture}
\caption{The essential Feynman diagrams that contribute at
$|q^2|,\ |u| \gg m^2$ in the collinear kinematics. Every set
of diagrams is gauge invariant relative to the heavy photon.
The signs before sets are defined by the Fermi statistics for
permutation of the final fermion states.}
\end{figure}
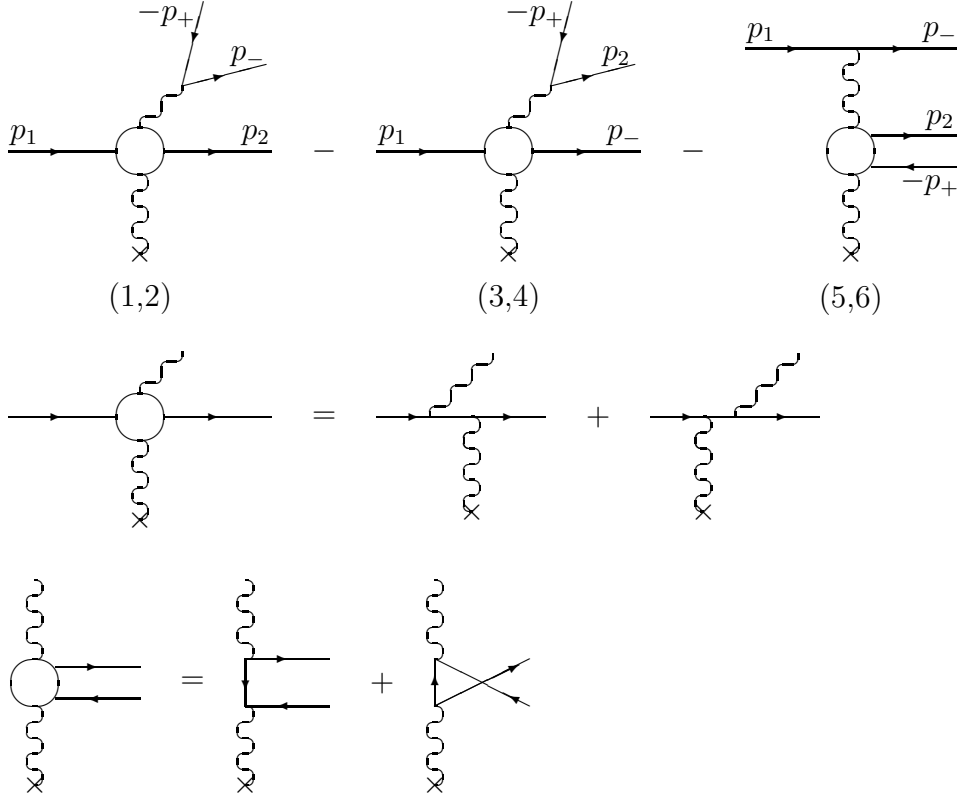

%(Essential Feynman diagrams for the process (2))
The amplitudes which correspond to every set of diagrams: (1,2), (3,4)
and (5,6) on the Fig.1 are gauge invariant, therefore it is convenient to
not separate these pairs at calculation. We will refer to them as (1,2)
set and so on.

\subsection{The contribution of the region $a)$}

\hspace{0.6cm}In the region $a)$ the (1,2) and (5,6) sets contribute. We
will define the limiting angle $\theta_0$ in this region such as
\begin{equation}\label{3,definition}
\theta_1 \ , \theta_2 \leq \theta_0 \ , \ \
(\varepsilon\theta_0/m)^2 = z_0 \gg 1 \ , \end{equation} where $\theta_1
(\theta_2) = \widehat{\vec p_1\vec p_+} (\widehat{\vec p_1\vec p_-}$) and
$\varepsilon$ is the initial electron energy. For the kinematical
invarians which correspond to the small propagator denominators in
essential Feynman diagrams we introduce the folowing parametrization
\cite{NPM89} $$ a= \frac{(p_+ + p_-)^2}{m^2} =
\frac{1}{x_1x_2}[(x_1+x_2)^2 + x_1^2x_2^2(\vec n_1 - \vec n_2)^2] \ , \
a_1 = \frac{2(p_1p_+)}{m^2} = \frac{1}{x_1}(1+x_1^2 +x_1^2\vec n_1^2) \ ,
$$ \begin{equation}\label{4,invariants} a_2 = \frac{2(p_1p_-)}{m^2} =
\frac{1}{x_2}(1+x_2^2 +x_2^2\vec n_2^2) \ , \ \Delta =
\frac{(p_1-p_--p_+)^2-m^2}{m^2} = a - a_1 - a_2 \ , \end{equation} $$x_1 =
\frac{\varepsilon_+}{\varepsilon} \ , \ \ x_2 =
\frac{\varepsilon_-}{\varepsilon} \ , \ \ |n_i| =
\frac{\varepsilon\theta_0}{m}\ , \ \ q_a = yp_1 - p_2 \ , \ \ y =
1-x_1 - x_2 \ , $$
where $\varepsilon_+ \ (\varepsilon_-)$ is the created positron (electron)
energy, and $\vec n_1\ , \vec n_2$ are two--dimensional vectors
perpendicular to direction of the 3--momentum $\vec p_1.$

We will define the ECT as a product of the created electron--positron
pair phase space and the trace--tensor of the corresponding tensor diagram
(TD) given on Fig.2

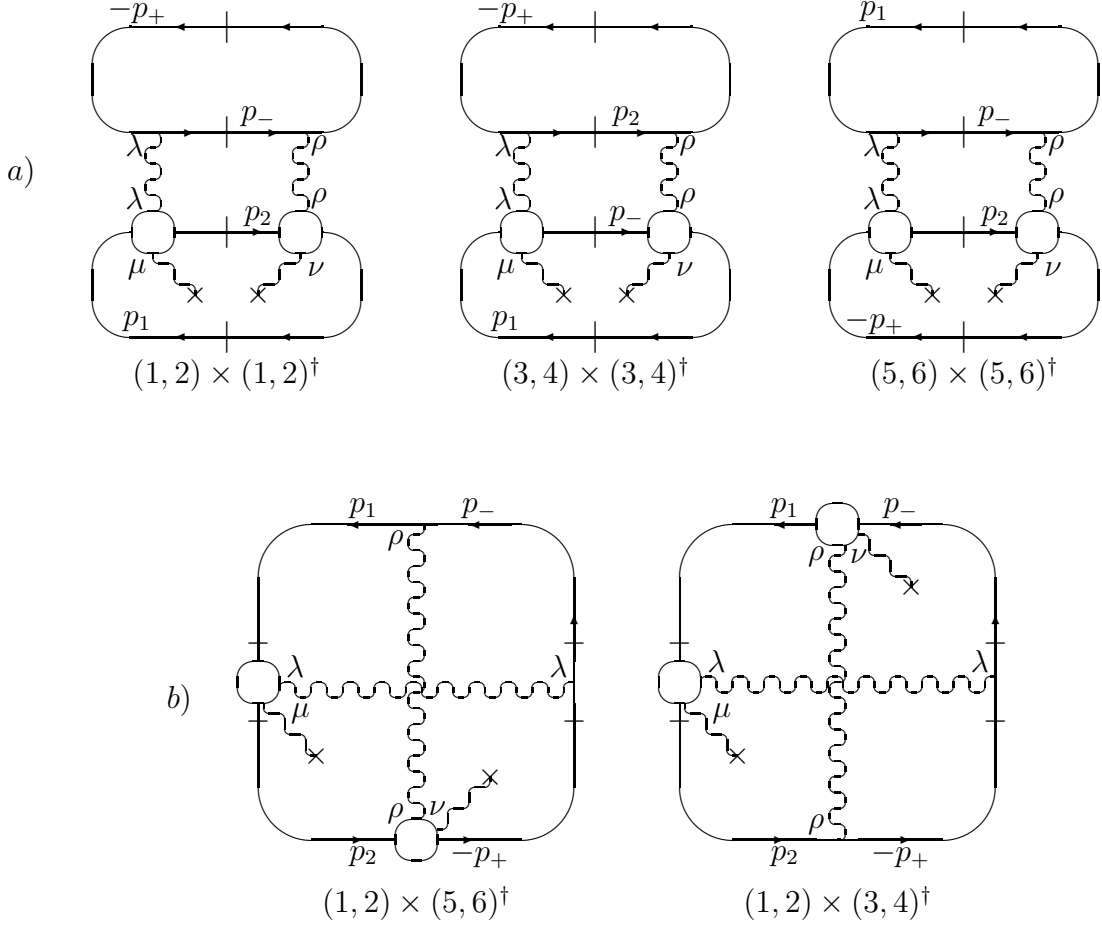
\begin{figure}[t]
\unitlength=0.70mm
\special{em:linewidth 0.6pt}
\linethickness{0.6pt}
\begin{picture}(190.00,80.00)
\put(05.00,41.50){\makebox(0,0)[cc]{$a)$}}
\put(44.00,03.00){\makebox(0,0)[cc]{$(1,2)\times(1,2)^{\dagger}$}}
\put(30.00,30.00){\oval(8.00,8.00)[]}
\put(34.00,30.00){\line(1,0){20.00}}
\put(34.00,30.00){\vector(1,0){17.00}}
\put(44.00,30.00){\makebox(0,0)[cc]{$|$}}
\put(26.00,20.00){\oval(15.00,20.00)[l]}
\put(26.00,10.00){\line(1,0){36.00}}
\put(58.00,30.00){\oval(8.00,8.00)[]}
\put(62.00,20.00){\oval(15.00,20.00)[r]}
\put(62.00,10.00){\vector(-1,0){8.00}}
\put(62.00,10.00){\vector(-1,0){28.00}}
\put(44.00,10.00){\makebox(0,0)[cc]{$|$}}
\put(30.00,35.50){\oval(3.00,3.00)[r]}
\put(30.00,38.50){\oval(3.00,3.00)[l]}
\put(30.00,41.50){\oval(3.00,3.00)[r]}
\put(30.00,44.50){\oval(3.00,3.00)[l]}
\put(30.00,47.50){\oval(3.00,3.00)[r]}
\put(26.50,36.50){\makebox(0,0)[cc]{$\lambda$}}
\put(26.50,46.50){\makebox(0,0)[cc]{$\lambda$}}
\put(58.00,35.50){\oval(3.00,3.00)[r]}
\put(58.00,38.50){\oval(3.00,3.00)[l]}
\put(58.00,41.50){\oval(3.00,3.00)[r]}
\put(58.00,44.50){\oval(3.00,3.00)[l]}
\put(58.00,47.50){\oval(3.00,3.00)[r]}
\put(61.50,36.50){\makebox(0,0)[cc]{$\rho$}}
\put(61.50,46.50){\makebox(0,0)[cc]{$\rho$}}
\put(50.00,33.00){\makebox(0,0)[cc]{$p_2$}}
\put(26.00,49.00){\line(1,0){36.00}}
\put(26.00,49.00){\vector(1,0){12.00}}
\put(26.00,49.00){\vector(1,0){28.00}}
\put(44.00,49.00){\makebox(0,0)[cc]{$|$}}
\put(26.00,59.00){\oval(15.00,20.00)[l]}
\put(62.00,59.00){\oval(15.00,20.00)[r]}
\put(26.00,69.00){\line(1,0){36.00}}
\put(62.00,69.00){\vector(-1,0){8.00}}
\put(62.00,69.00){\vector(-1,0){28.00}}
\put(44.00,69.00){\makebox(0,0)[cc]{$|$}}

\put(32.00,26.00){\oval(4.00,4.00)[lb]}
\put(32.00,22.00){\oval(4.00,4.00)[rt]}
\put(36.00,22.00){\oval(4.00,4.00)[lb]}
\put(36.00,18.00){\oval(4.00,4.00)[rt]}
\put(38.00,18.00){\makebox(0,0)[cc]{$\times$}}
\put(27.00,23.00){\makebox(0,0)[cc]{$\mu$}}
\put(27.00,13.00){\makebox(0,0)[cc]{$p_1$}}
\put(56.00,26.00){\oval(4.00,4.00)[rb]}
\put(56.00,22.00){\oval(4.00,4.00)[lt]}
\put(52.00,22.00){\oval(4.00,4.00)[rb]}
\put(52.00,18.00){\oval(4.00,4.00)[lt]}
\put(50.00,18.00){\makebox(0,0)[cc]{$\times$}}
\put(61.00,23.00){\makebox(0,0)[cc]{$\nu$}}
\put(50.00,52.00){\makebox(0,0)[cc]{$p_-$}}
\put(27.00,72.00){\makebox(0,0)[cc]{$-p_+$}}
%-----------------------------------------------------
\put(114.00,03.00){\makebox(0,0)[cc]{$(3,4)\times(3,4)^{\dagger}$}}
\put(100.00,30.00){\oval(8.00,8.00)[]}
\put(104.00,30.00){\line(1,0){20.00}}
\put(104.00,30.00){\vector(1,0){17.00}}
\put(114.00,30.00){\makebox(0,0)[cc]{$|$}}
\put(96.00,20.00){\oval(15.00,20.00)[l]}
\put(96.00,10.00){\line(1,0){36.00}}
\put(128.00,30.00){\oval(8.00,8.00)[]}
\put(132.00,20.00){\oval(15.00,20.00)[r]}
\put(132.00,10.00){\vector(-1,0){8.00}}
\put(132.00,10.00){\vector(-1,0){28.00}}
\put(114.00,10.00){\makebox(0,0)[cc]{$|$}}
\put(100.00,35.50){\oval(3.00,3.00)[r]}
\put(100.00,38.50){\oval(3.00,3.00)[l]}
\put(100.00,41.50){\oval(3.00,3.00)[r]}
\put(100.00,44.50){\oval(3.00,3.00)[l]}
\put(100.00,47.50){\oval(3.00,3.00)[r]}
\put(96.50,36.50){\makebox(0,0)[cc]{$\lambda$}}
\put(96.50,46.50){\makebox(0,0)[cc]{$\lambda$}}
\put(128.00,35.50){\oval(3.00,3.00)[r]}
\put(128.00,38.50){\oval(3.00,3.00)[l]}
\put(128.00,41.50){\oval(3.00,3.00)[r]}
\put(128.00,44.50){\oval(3.00,3.00)[l]}
\put(128.00,47.50){\oval(3.00,3.00)[r]}
\put(131.50,36.50){\makebox(0,0)[cc]{$\rho$}}
\put(131.50,46.50){\makebox(0,0)[cc]{$\rho$}}
\put(120.00,33.00){\makebox(0,0)[cc]{$p_-$}}
\put(96.00,49.00){\line(1,0){36.00}}
\put(96.00,49.00){\vector(1,0){12.00}}
\put(96.00,49.00){\vector(1,0){28.00}}
\put(114.00,49.00){\makebox(0,0)[cc]{$|$}}
\put(96.00,59.00){\oval(15.00,20.00)[l]}
\put(132.00,59.00){\oval(15.00,20.00)[r]}
\put(96.00,69.00){\line(1,0){36.00}}
\put(132.00,69.00){\vector(-1,0){8.00}}
\put(132.00,69.00){\vector(-1,0){28.00}}
\put(114.00,69.00){\makebox(0,0)[cc]{$|$}}
\put(102.00,26.00){\oval(4.00,4.00)[lb]}
\put(102.00,22.00){\oval(4.00,4.00)[rt]}
\put(106.00,22.00){\oval(4.00,4.00)[lb]}
\put(106.00,18.00){\oval(4.00,4.00)[rt]}
\put(108.00,18.00){\makebox(0,0)[cc]{$\times$}}
\put(97.00,23.00){\makebox(0,0)[cc]{$\mu$}}
\put(97.00,13.00){\makebox(0,0)[cc]{$p_1$}}
\put(126.00,26.00){\oval(4.00,4.00)[rb]}
\put(126.00,22.00){\oval(4.00,4.00)[lt]}
\put(122.00,22.00){\oval(4.00,4.00)[rb]}
\put(122.00,18.00){\oval(4.00,4.00)[lt]}
\put(120.00,18.00){\makebox(0,0)[cc]{$\times$}}
\put(131.00,23.00){\makebox(0,0)[cc]{$\nu$}}
\put(120.00,52.00){\makebox(0,0)[cc]{$p_2$}}
\put(97.00,72.00){\makebox(0,0)[cc]{$-p_+$}}
%-----------------------------------------------------
\put(184.00,03.00){\makebox(0,0)[cc]{$(5,6)\times(5,6)^{\dagger}$}}
\put(170.00,30.00){\oval(8.00,8.00)[]}
\put(174.00,30.00){\line(1,0){20.00}}
\put(174.00,30.00){\vector(1,0){17.00}}
\put(184.00,30.00){\makebox(0,0)[cc]{$|$}}
\put(166.00,20.00){\oval(15.00,20.00)[l]}
\put(166.00,10.00){\line(1,0){36.00}}
\put(198.00,30.00){\oval(8.00,8.00)[]}
\put(202.00,20.00){\oval(15.00,20.00)[r]}
\put(202.00,10.00){\vector(-1,0){8.00}}
\put(202.00,10.00){\vector(-1,0){28.00}}
\put(184.00,10.00){\makebox(0,0)[cc]{$|$}}
\put(170.00,35.50){\oval(3.00,3.00)[r]}
\put(170.00,38.50){\oval(3.00,3.00)[l]}
\put(170.00,41.50){\oval(3.00,3.00)[r]}
\put(170.00,44.50){\oval(3.00,3.00)[l]}
\put(170.00,47.50){\oval(3.00,3.00)[r]}
\put(166.50,36.50){\makebox(0,0)[cc]{$\lambda$}}
\put(166.50,46.50){\makebox(0,0)[cc]{$\lambda$}}
\put(198.00,35.50){\oval(3.00,3.00)[r]}
\put(198.00,38.50){\oval(3.00,3.00)[l]}
\put(198.00,41.50){\oval(3.00,3.00)[r]}
\put(198.00,44.50){\oval(3.00,3.00)[l]}
\put(198.00,47.50){\oval(3.00,3.00)[r]}
\put(201.50,36.50){\makebox(0,0)[cc]{$\rho$}}
\put(201.50,46.50){\makebox(0,0)[cc]{$\rho$}}
\put(190.00,33.00){\makebox(0,0)[cc]{$p_2$}}
\put(166.00,49.00){\line(1,0){36.00}}
\put(166.00,49.00){\vector(1,0){12.00}}
\put(166.00,49.00){\vector(1,0){28.00}}
\put(184.00,49.00){\makebox(0,0)[cc]{$|$}}
\put(166.00,59.00){\oval(15.00,20.00)[l]}
\put(202.00,59.00){\oval(15.00,20.00)[r]}
\put(166.00,69.00){\line(1,0){36.00}}
\put(202.00,69.00){\vector(-1,0){8.00}}
\put(202.00,69.00){\vector(-1,0){28.00}}
\put(184.00,69.00){\makebox(0,0)[cc]{$|$}}
\put(172.00,26.00){\oval(4.00,4.00)[lb]}
\put(172.00,22.00){\oval(4.00,4.00)[rt]}
\put(176.00,22.00){\oval(4.00,4.00)[lb]}
\put(176.00,18.00){\oval(4.00,4.00)[rt]}
\put(178.00,18.00){\makebox(0,0)[cc]{$\times$}}
\put(167.00,23.00){\makebox(0,0)[cc]{$\mu$}}
\put(167.00,13.00){\makebox(0,0)[cc]{$-p_+$}}
\put(196.00,26.00){\oval(4.00,4.00)[rb]}
\put(196.00,22.00){\oval(4.00,4.00)[lt]}
\put(192.00,22.00){\oval(4.00,4.00)[rb]}
\put(192.00,18.00){\oval(4.00,4.00)[lt]}
\put(190.00,18.00){\makebox(0,0)[cc]{$\times$}}
\put(201.00,23.00){\makebox(0,0)[cc]{$\nu$}}
\put(190.00,52.00){\makebox(0,0)[cc]{$p_-$}}
\put(167.00,72.00){\makebox(0,0)[cc]{$p_1$}}
\end{picture}

\begin{picture}(190.00,100.00)
\put(35.00,41.50){\makebox(0,0)[cc]{$b)$}}
\put(80.00,03.00){\makebox(0,0)[cc]{$(1,2)\times(5,6)^{\dagger}$}}
\put(80.00,15.00){\oval(8.00,8.00)[]}
\put(50.00,45.00){\oval(8.00,8.00)[]}
\put(76.00,41.00){\oval(52.00,52.00)[lb]}
\put(84.00,41.00){\oval(52.00,52.00)[rb]}
\put(76.00,49.00){\oval(52.00,52.00)[lt]}
\put(84.00,49.00){\oval(52.00,52.00)[rt]}
\put(84.00,75.00){\vector(-1,0){17.00}}
\put(98.00,75.00){\vector(-1,0){08.00}}
\put(110.00,41.00){\vector(0,1){17.00}}
\put(84.00,15.00){\vector(1,0){7.00}}
\put(64.00,15.00){\vector(1,0){6.00}}
\put(80.00,20.50){\oval(3.00,3.00)[r]}
\put(80.00,23.50){\oval(3.00,3.00)[l]}
\put(80.00,26.50){\oval(3.00,3.00)[r]}
\put(80.00,29.50){\oval(3.00,3.00)[l]}
\put(80.00,32.50){\oval(3.00,3.00)[r]}
\put(80.00,35.50){\oval(3.00,3.00)[l]}
%\put(80.00,41.50){\oval(3.00,9.00)[r]}
\put(80.00,38.50){\oval(3.00,3.00)[r]}
\put(80.00,41.50){\oval(3.00,3.00)[l]}
\put(80.00,44.50){\oval(3.00,3.00)[r]}
\put(80.00,47.50){\oval(3.00,3.00)[l]}
\put(80.00,50.50){\oval(3.00,3.00)[r]}
\put(80.00,53.50){\oval(3.00,3.00)[l]}
\put(80.00,56.50){\oval(3.00,3.00)[r]}
\put(80.00,59.50){\oval(3.00,3.00)[l]}
\put(80.00,62.50){\oval(3.00,3.00)[r]}
\put(80.00,65.50){\oval(3.00,3.00)[l]}
\put(80.00,68.50){\oval(3.00,3.00)[r]}
\put(80.00,71.50){\oval(3.00,3.00)[l]}
\put(80.00,75.00){\oval(3.00,4.00)[rb]}
\put(55.50,43.50){\oval(3.00,3.00)[t]}
\put(58.50,43.50){\oval(3.00,3.00)[b]}
\put(61.50,43.50){\oval(3.00,3.00)[t]}
\put(64.50,43.50){\oval(3.00,3.00)[b]}
\put(67.50,43.50){\oval(3.00,3.00)[t]}
\put(70.50,43.50){\oval(3.00,3.00)[b]}
\put(73.50,43.50){\oval(3.00,3.00)[t]}
\put(76.50,43.50){\oval(3.00,3.00)[b]}
\put(79.50,43.50){\oval(3.00,3.00)[t]}
\put(82.50,43.50){\oval(3.00,3.00)[b]}
%\put(79.50,43.50){\oval(9.00,3.00)[b]}
\put(85.50,43.50){\oval(3.00,3.00)[t]}
\put(88.50,43.50){\oval(3.00,3.00)[b]}
\put(91.50,43.50){\oval(3.00,3.00)[t]}
\put(94.50,43.50){\oval(3.00,3.00)[b]}
\put(97.50,43.50){\oval(3.00,3.00)[t]}
\put(100.50,43.50){\oval(3.00,3.00)[b]}
\put(103.50,43.50){\oval(3.00,3.00)[t]}
\put(106.50,43.50){\oval(3.00,3.00)[b]}
\put(110.00,43.50){\oval(4.00,3.00)[lt]}
\put(53.00,41.00){\oval(4.00,4.00)[lb]}
\put(53.00,37.00){\oval(4.00,4.00)[rt]}
\put(57.00,37.00){\oval(4.00,4.00)[lb]}
\put(57.00,33.00){\oval(4.00,4.00)[rt]}
\put(61.00,33.00){\oval(4.00,4.00)[lb]}
\put(61.00,31.00){\makebox(0,0)[cc]{$\times$}}
\put(84.00,19.00){\oval(4.00,4.00)[rb]}
\put(88.00,19.00){\oval(4.00,4.00)[lt]}
\put(88.00,23.00){\oval(4.00,4.00)[rb]}
\put(92.00,23.00){\oval(4.00,4.00)[lt]}
\put(92.00,27.00){\oval(4.00,4.00)[rb]}
\put(94.00,27.00){\makebox(0,0)[cc]{$\times$}}
\put(84.00,21.00){\makebox(0,0)[cc]{$\nu$}}
\put(76.00,20.00){\makebox(0,0)[cc]{$\rho$}}
\put(76.00,72.00){\makebox(0,0)[cc]{$\rho$}}
\put(58.00,39.00){\makebox(0,0)[cc]{$\mu$}}
\put(57.00,48.00){\makebox(0,0)[cc]{$\lambda$}}
\put(107.00,48.00){\makebox(0,0)[cc]{$\lambda$}}
\put(50.00,37.50){\makebox(0,0)[cc]{$-$}}
\put(50.00,52.50){\makebox(0,0)[cc]{$-$}}
\put(110.00,37.50){\makebox(0,0)[cc]{$-$}}
\put(110.00,52.50){\makebox(0,0)[cc]{$-$}}
\put(70.00,12.00){\makebox(0,0)[cc]{$p_2$}}
\put(92.00,12.00){\makebox(0,0)[cc]{$-p_+$}}
\put(70.00,78.00){\makebox(0,0)[cc]{$p_1$}}
\put(92.00,78.00){\makebox(0,0)[cc]{$p_-$}}
%---------------------------------------------------
\put(160.00,03.00){\makebox(0,0)[cc]{$(1,2)\times(3,4)^{\dagger}$}}
\put(160.00,75.00){\oval(8.00,8.00)[]}
\put(130.00,45.00){\oval(8.00,8.00)[]}
\put(156.00,41.00){\oval(52.00,52.00)[lb]}
\put(164.00,41.00){\oval(52.00,52.00)[rb]}
\put(156.00,49.00){\oval(52.00,52.00)[lt]}
\put(164.00,49.00){\oval(52.00,52.00)[rt]}
\put(150.00,15.00){\vector(1,0){24.00}}
\put(142.00,15.00){\vector(1,0){08.00}}
\put(190.00,41.00){\vector(0,1){17.00}}
\put(156.00,75.00){\vector(-1,0){8.00}}
\put(176.00,75.00){\vector(-1,0){6.00}}
\put(160.00,16.50){\oval(3.00,3.00)[r]}
\put(160.00,19.50){\oval(3.00,3.00)[l]}
\put(160.00,22.50){\oval(3.00,3.00)[r]}
\put(160.00,25.50){\oval(3.00,3.00)[l]}
\put(160.00,28.50){\oval(3.00,3.00)[r]}
\put(160.00,31.50){\oval(3.00,3.00)[l]}
%\put(160.00,37.50){\oval(3.00,9.00)[r]}
\put(160.00,34.50){\oval(3.00,3.00)[r]}
\put(160.00,37.50){\oval(3.00,3.00)[l]}
\put(160.00,40.50){\oval(3.00,3.00)[r]}
\put(160.00,43.50){\oval(3.00,3.00)[l]}
\put(160.00,46.50){\oval(3.00,3.00)[r]}
\put(160.00,49.50){\oval(3.00,3.00)[l]}
\put(160.00,52.50){\oval(3.00,3.00)[r]}
\put(160.00,55.50){\oval(3.00,3.00)[l]}
\put(160.00,58.50){\oval(3.00,3.00)[r]}
\put(160.00,61.50){\oval(3.00,3.00)[l]}
\put(160.00,64.50){\oval(3.00,3.00)[r]}
\put(160.00,67.50){\oval(3.00,3.00)[l]}
\put(160.00,71.00){\oval(3.00,4.00)[rb]}
\put(135.50,44.50){\oval(3.00,3.00)[t]}
\put(138.50,44.50){\oval(3.00,3.00)[b]}
\put(141.50,44.50){\oval(3.00,3.00)[t]}
\put(144.50,44.50){\oval(3.00,3.00)[b]}
\put(147.50,44.50){\oval(3.00,3.00)[t]}
\put(150.50,44.50){\oval(3.00,3.00)[b]}
\put(153.50,44.50){\oval(3.00,3.00)[t]}
\put(156.50,44.50){\oval(3.00,3.00)[b]}
\put(159.50,44.50){\oval(3.00,3.00)[t]}
\put(162.50,44.50){\oval(3.00,3.00)[b]}
%\put(159.50,44.50){\oval(9.00,3.00)[b]}
\put(165.50,44.50){\oval(3.00,3.00)[t]}
\put(168.50,44.50){\oval(3.00,3.00)[b]}
\put(171.50,44.50){\oval(3.00,3.00)[t]}
\put(174.50,44.50){\oval(3.00,3.00)[b]}
\put(177.50,44.50){\oval(3.00,3.00)[t]}
\put(180.50,44.50){\oval(3.00,3.00)[b]}
\put(183.50,44.50){\oval(3.00,3.00)[t]}
\put(186.50,44.50){\oval(3.00,3.00)[b]}
\put(190.00,44.50){\oval(4.00,3.00)[lt]}
\put(133.00,41.00){\oval(4.00,4.00)[lb]}
\put(133.00,37.00){\oval(4.00,4.00)[rt]}
\put(137.00,37.00){\oval(4.00,4.00)[lb]}
\put(137.00,33.00){\oval(4.00,4.00)[rt]}
\put(141.00,33.00){\oval(4.00,4.00)[lb]}
\put(141.00,31.00){\makebox(0,0)[cc]{$\times$}}
\put(164.00,71.00){\oval(4.00,4.00)[rt]}
\put(168.00,71.00){\oval(4.00,4.00)[lb]}
\put(168.00,67.00){\oval(4.00,4.00)[rt]}
\put(172.00,67.00){\oval(4.00,4.00)[lb]}
\put(172.00,63.00){\oval(4.00,4.00)[rt]}
\put(174.00,63.00){\makebox(0,0)[cc]{$\times$}}
\put(164.00,69.00){\makebox(0,0)[cc]{$\nu$}}
\put(156.00,18.00){\makebox(0,0)[cc]{$\rho$}}
\put(156.00,69.00){\makebox(0,0)[cc]{$\rho$}}
\put(138.00,39.00){\makebox(0,0)[cc]{$\mu$}}
\put(137.00,49.00){\makebox(0,0)[cc]{$\lambda$}}
\put(187.00,49.00){\makebox(0,0)[cc]{$\lambda$}}
\put(130.00,37.50){\makebox(0,0)[cc]{$-$}}
\put(130.00,52.50){\makebox(0,0)[cc]{$-$}}
\put(190.00,37.50){\makebox(0,0)[cc]{$-$}}
\put(190.00,52.50){\makebox(0,0)[cc]{$-$}}
\put(150.00,12.00){\makebox(0,0)[cc]{$p_2$}}
\put(172.00,12.00){\makebox(0,0)[cc]{$-p_+$}}
\put(150.00,78.00){\makebox(0,0)[cc]{$p_1$}}
\put(172.00,78.00){\makebox(0,0)[cc]{$p_-$}}
\end{picture}
\caption{Diagrams for the electron current tensor (TD). Topological
class $a)$ is responsible for the contributions of (1,2), (3,4),
and (5,6) sets, that appear in both collinear and semicollinear
kinematical regions. Topological class $b)$ describes the
interference between (1,2) set and (3,4) as well as (5,6) ones.
The latter class contributes in collinear kinematics only within
the chosen accuracy.}
\end{figure}

\begin{equation}\label{5,definition of ECT}
L_{\mu\nu} =
\bigl(\frac{\alpha}{4\pi^2}\bigr)^2\frac{d^3p_+d^3p_-}{\varepsilon_+
\varepsilon_-}T_{\mu\nu} \ ,
\end{equation}
where the starting--points for calculation of the tensor $T_{\mu\nu}$ are
defined in Appendix in terms of the corresponding matrix element squared.
In the region $a)$ the created pair phase space can be written by means of
the used in (3) variables as follows \begin{equation}\label{6,phase space}
\frac{d^3p_+d^3p_-}{\varepsilon_+\varepsilon_-} = m^4\pi^2x_1x_2dx_1dx_2
dz_1dz_2\frac{d\phi}{2\pi}\ , \ \ z_{1,2} =
\frac{\varepsilon^2\theta_{1,2}^2}{m^2} \ ,
\end{equation}
where $\phi $ is the angle between vectors $\vec n_1$ and $\vec n_2.$

When calculating tensor $T_{\mu\nu}$ we have to leave terms with $m^4$
in the denominator and neglect with terms of the order $m^{-2}.$ Such
approach allows, in principle, to compute the quantity $L_{\mu\nu}$ with
the power accuracy relative parameter $z_0$ neglecting only with terms of
the order $O(z_0^{-1}).$

In the region $a)$ it is convenient to separate the contributions of (1,2)
and (5,6) sets and their interference
$$ T_{\mu\nu}^{^{a)}} = \frac{16}{m^4}[T_{\mu\nu}^{(12)} +
T_{\mu\nu}^{(56)} + T_{\mu\nu}^{(int)}] \ . $$ The tensor
$T_{\mu\nu}^{(12)}$ can be written as follows
\begin{equation}\label{7,(1,2) set} T_{\mu\nu}^{(12)} =
\bigl\{-\frac{1}{a^2\Delta^2}\bigl[\frac{2y}{(1-y)^2}(x_2a_1-x_1a_2)^2 +
4ay + 2(1-y)\Delta\bigr] + \frac{2x_1x_2}{\Delta^2(1-y)^2} -
\frac{1+y}{a\Delta(1-y)}
\end{equation}
$$-\frac{1}{a\Delta^2(1-y)^2}[y(1-y)(x_1-x_2)(a_1-a_2) +
2x_1x_2(1+y)(a_1+a_2)]\bigr\}(Q_{\mu\nu} + i\lambda E_{\mu\nu})$$
$$-\frac{i\lambda}{a^2\Delta^2}[2a((1-y)^2-2x_1x_2)+4(1-y)^2]E_{\mu\nu} \
.  $$ It needs to emphasize that the last term on the right side of Eq.(7)
does not give the large logarithm being integrated over the
angular phase space of the created pair. Therefore, in the frame of
leading and next--to--leading accuracy both tensor structure, $Q_{\mu\nu}$
and $i\lambda E_{\mu\nu},$ are multiplied by the same function.
Cosequently, with the such accuracy here we can use the result of the
corresponding calculations for unpolarized case [8,9] where only
$Q_{\mu\nu}$ structure appears.

As concerns the tensor $T_{\mu\nu}^{^{(5,6)}},$ its symmetrical
spin--independent part can be obtained from the corresponding part of
$T_{\mu\nu}$ by the rule
\begin{equation}\label{8,rule for (1,2)to(5,6)}
sym\{T_{\mu\nu}^{(56)}\} = -x_1\hat P sym\{T_{\mu\nu}^{(12)}\} \ ,
\end{equation}
where the substitution operator $\hat P$ acts as follows
\begin{equation}\label{9,action of operator P}
\hat P (x_1, \ x_2 \ , \vec n_1 \ , \vec n_2) \rightarrow ( x_1, \
-\frac{x_2}{x_1} \ , \ x_1\vec n_1\ , \ \ x_1(\vec n_1 - \vec n_2) )\ .
\end{equation}
It easy to verify that
$$ \hat P (a_1\ , \ a \ , \ a_2) \rightarrow (a_1\ , \ 2-a_2\ , 2-a) \ .$$
The rule (9) reflects the obvious topological equivalence ot TD for (1,2)
and (5,6) sets in unpolarized case as it follows from the Fig.2a (see also
\cite{NPMU89}).

Unfortunately, we cannot use this rule to obtain the antisymmetrical
spin--dependent part of the tensor $T_{\mu\nu}^{(56)}.$ On the level of TD
of Fig.2 we can explain this fact because for the (1,2) set the polarised
particle enters to the lower block while for the (5,6) set -- to the upper
one. It is obvious that conditions for the polarized particle are
different in lower and upper blocks.

The strightforward caculations leads to the following expression
\begin{equation}\label{10, tensor for (5,6) set}
T_{\mu\nu}^{(56)} = S^{(56)}Q_{\mu\nu} + A^{(56)}i\lambda E_{\mu\nu} \ ,
\end{equation}
where
\begin{equation}\label{11, sym (5,6)}
S^{(56)} = -\frac{2y}{(2-a_2)^2}\bigl(\frac{a_1}{\Delta}+\frac{1}{1-x_2}
\bigr)^2 - \frac{1}{\Delta(2-a_2)}\bigl(x_1-y-\frac{2(1-x_1+x_1^2)}
{1-x_2}\bigr) - \frac{y}{\Delta^2} + \frac{2ya_1}{\Delta^2(2-a_2)}
\end{equation}
$$+ \frac{2(1-x_2)}{\Delta(2-a_2)^2} - \frac{4y}{\Delta^2(2-a_2)}\ , $$
\begin{equation}\label{12,asym (5,6)}
 A^{(56)} = -\frac{1}{\Delta(2-a_2)(1-x_2)}(1-x_2+x_2(x_1-y)) +
 \frac{y}{\Delta^2} -\frac{2ya_1}{\Delta^2(2-a_2)} -
\frac{2(1-x_2)(x_1-y)}{\Delta(2-a_2)^2} \end{equation}
$$-\frac{2[2(1-x_1)-(1-y)(1-x_2)]}{\Delta^2(2-a_2)} -
\frac{4(1-x_2)^2}{\Delta^2(2-a_2)^2} \ . $$
We see that in contrast with the contribution of (1,2) set, for which the
structures $Q_{\mu\nu}$ and $i\lambda E_{\mu\nu}$ are accompanied by the
same function up to next--to--leading accuracy, quantities $S^{(56)}$ and
$A^{(56)}$ on the right side of Eq.(10) are quite different. As we will
see below this difference affects yet in leading approximation.

The interference of (1,2) and (5,6) sets (see TD Fig.2b) can be written as
follows \begin{equation}\label{13,int (5,6)} T_{\mu\nu}^{^{(int)}} =
S^{^{(int)}}( Q_{\mu\nu} + i\lambda E_{\mu\nu}) + A^{^{(int)}}i\lambda
E_{\mu\nu} \ , \end{equation}
$$S^{^{(int)}} = - \frac{1}{a\Delta}\bigl[2x_2 +
(yx_1-x_2)\bigl(\frac{1}{1-x_2}+\frac{1}{1-y}\bigr)\bigr] -
\frac{1}{\Delta(2-a_2)}\bigl[2x_2  $$
$$+(y-x_1x_2)\bigl(\frac{1}{1-x_2}+\frac{1}{1-y}\bigr)\bigr]
- \frac{1}{a(2-a_2)}\bigl[-\frac{2x_1x_2}{(1-x_2)(1-y)} $$
\begin{equation}\label{14, sym int (56)}
+ x_2(1+x_1)
\bigl(\frac{1}{1-x_2}+\frac{1}{1-y}\bigr) -2x_2\bigr] -
\frac{2ya_1}{\Delta^2}\bigl(\frac{1}{a}+\frac{1}{2-a_2}\bigr)
\end{equation}
$$-\frac{2(yx_2-x_1)}{\Delta a(2-a_2)}\bigl(\frac{1}{1-x_2}-\frac{1}{1-y}
\bigr) - \frac{4y}{\Delta^2}\bigl(\frac{1}{a}+\frac{1}{2-a_2}\bigr)
+ \frac{8y}{a\Delta^2(2-a_2)} \ , $$
\begin{equation}\label{15, asym int (56)}
A^{^{(int)}} = -\frac{2(1-y)}{a\Delta(2-a_2)}(2x_1-x_2+3x_2^2) -
\frac{2}{\Delta^2(2-a_2)}(x_1-x_2+2x_1^2+3x_2(1-y))
\end{equation}
$$+\frac{2(1-y)^2}{a\Delta^2}\bigl(3-\frac{4}{2-a_2}\bigr) \ .$$
The quantity $S^{^{(int)}}$ is invariant relative the same operation which
transforms tensor $T_{\mu\nu}^{^{(12)}}$ into tensor $T_{\mu\nu}^{^{(56)}}$
\begin{equation}\label{16, inariance}
S^{^{(int)}} = -x_1\hat PS^{^{(int)}} \ ,
\end{equation}
while the quantity $A^{^{(int)}}$ is not invariant.

The last line on the right side of Eq.(14) as well as $A^{^{(int)}}$ does
not contribute in the frame of logarithic accuracy.

The next step in our calculations is the integration of the ECT over
angular phase space of the created pair. We use the parametrization (6)
and perform the integration over variables $z_1$ and $z_2$ from $0$ up to
$z_0$ and over $\phi$ -- from $0$ up to $2\pi.$ In principle, the angular
integration can be done with power accuracy, but in this article we
restrict ourselves with next--to--leading one. The method of
integration suitable for the such approximation is described in
\cite{NPMphot88} and the table of corresponding integrals is given in
[8,11]. As we noted above (see Eqs.(7),(13)) at such accuracy the
contribution in ECT due to (1,2) set and interference of (1,2) and (5,6)
sets on differential level contains the same function at symmetrical and
antisymmetrical structures.  Therefore, we can write the result of angular
integration in the region $a)$ in the form \begin{equation}\label{17,
integrated form} \overline{L_{\mu\nu}^{^{a)}}} =
\frac{\alpha^2}{\pi^2}dx_1dx_2\ln z_0[(Q_{\mu\nu} + i\lambda
E_{\mu\nu})\overline{S} + \overline{S^{^{(56)}}}Q_{\mu\nu} + i\lambda
\overline{A^{^{(56)}}}E_{\mu\nu}] \ , \end{equation} where $\overline{S}$
absorbs the results of angular integration of the right sides of Eqs.(7)
and (14) \begin{equation}\label{18, (12) and int} \overline{S} =
\frac{1+y^2}{(1-y)^4}\bigl[\frac{1}{2}(x_1^2+x_2^2)\ln\frac{z_0x_1^2x_2^2}
{y^2} - (x_1-x_2)^2 + \frac{8yx_1x_2}{1+y^2}\bigr]
\end{equation}
$$+2\bigl[-\frac{x_2^2+y^2}{(1-x_2)(1-y)}\ln\frac{(1-y)(1-x_2)}{yx_2} +
\frac{x_1x_2-y}{(1-x_2)^2} + \frac{yx_1-x_2}{(1-y)^2}\bigr] \ . $$
The symmetrical part of the contribution connected with (5,6) set is
defined by the formula
\begin{equation}\label{19, sym (56)}
\overline{S^{^{(56)}}} =
\frac{x_1^2+y^2}{(1-x_2)^4}\bigl[\frac{1}{2}(1+x_2^2)\ln
\frac{z_0x_1^2x_2^2}{y^2} - (1+x_2)^2 + \frac{8yx_1x_2}{x_1^2+y^2}\bigr]
\ ,
\end{equation}
and the corresponding antisymmetrical part reads
\begin{equation}\label{20,asym (56)}
\overline{A^{^{(56)}}} =
\frac{(1+x_2)(y-x_1)}{2(1-x_2)^2}\ln\frac{z_0x_1^2x_2^2}{y^2} - 3 +
\frac{4+6y}{1-x_2} - \frac{8y}{(1-x_2)^2}\ .
\end{equation}
Note that coefficient at $Q_{\mu\nu}$ structure on the right side of
Eq.(17) is invariant relative substitution
$$ (x_1 \ , x_2 \ , z_0) \rightarrow (\frac{1}{x_1}\ , -\frac{x_2}{x_1}\ ,
x_1^2z_0) \ , $$
and that is the consequence of relations (8) and (16).

Because in the region $a)$ the heavy photon 4--momentum $q_a$ depends on
the sum of the created electron and positron energy fractions (see (4)) we
can integrate the right side of Eq.(17) over the electron (or positron)
energy fraction $x_2$ (or $x_1$) at fixed $x_1+x_2=1-y .$ The result can
be written as follows
\begin{equation}\label{21, final result for a)}
\widetilde L_{\mu\nu}^{^{a)}} = \frac{\alpha^2}{\pi^2}dy\ln
z_0[(Q_{\mu\nu}+i\lambda E_{\mu\nu})\widetilde S + \widetilde S^{^{(56)}}
Q_{\mu\nu} + i\lambda\widetilde A^{^{(56)}}E_{\mu\nu}] \ ,
\end{equation}
where
\begin{equation}\label{22}
\widetilde S = \frac{1+y^2}{3(1-y)}\ln z_0 + 4(1+y)\ln y\ln(1-y) +
\frac{2(3y^2-1)}{1-y}L_{i2}(1-y) - \frac{2}{1-y}\ln^2y
\end{equation}
$$+ \frac{8}{3(1-y)}\ln(1-y)-\frac{20}{9(1-y)} \ ,$$
\begin{equation}\label{23}
\widetilde S^{^{(56)}} = \bigl[\frac{1-y}{6y}(4+7y+4y^2) + (1+y)\ln
y\bigr]\ln z_0 + \frac{2}{3}\bigl(-4y^2-5y+1+\frac{4}{y}\bigr)\ln(1-y)\ ,
\end{equation}
$$+\frac{1}{3}\bigl(8y^2+5y-7-\frac{13}{1-y}\bigr)\ln y - \frac{2}{3}y^2
+ \frac{136}{9}y - \frac{107}{9} - \frac{4}{3y} \ , $$
\begin{equation}\label{24}
\widetilde A^{^{(56)}} = \frac{1}{2}[5(1-y)+2(1+y)\ln y]\ln z_0 +
\frac{2}{3}(13-17y)\ln(1-y) + \frac{1}{3}\bigl(5y-19-\frac{13}{1-y}\bigr)
\ln y
\end{equation}
$$+\frac{196}{9}y - \frac{185}{9}\ .$$

Let us pay attention on the leading (double logarithmic) contribution into
tensor $\widetilde L_{\mu\nu}^{^{a)}}$ ( the first terms on the right
sides of Eqs.(22),(23) and (24)). Terms which enter in Eqs.(22) and (23)
are well known as the electron structure function due to pair
production \cite{ESF}.

The first one is resposible for nonsinglet channel contribution. It has
infrared singularity at $y\rightarrow 1$ and can be obtained by insertion
of effective electromagnetic coupling (which is integral of running
constant) into so--called $\Theta$--term of the first--order electron
structure function \cite{ESF}. Thus, we see that in nonsinglet
channel the spin--independent part of the ECT and spin--dependent one have
the same behaviour in the leading approximation. This is true for pair
production as well as for photon emission \cite{KM98} and is the consequence of
the helicity conservation in the nosinglet channel \cite{BKL83}.

The second one describes the spin--independent part of the
ECT in singlet channel. It has a specific $ y^{-1}$ behaviour at
small values of $y$ and goes to zero when $y\rightarrow 1.$ But the
corresponding spin--dependent part of ECT in Eq.(24) is described by quite
another structure function
\begin{equation}\label{25, leading spin-dependent part}
\widetilde A_L^{^{(56)}}(y) = \frac{1}{2}[5(1-y)+2(1+y)\ln y]\ln^2z_0 \ ,
\ \ \int\limits_0^1\widetilde A_L^{^{(56)}}(y)dy = 0 \ ,
\end{equation}
which as well goes to zero at $y\rightarrow 1$ but has quite different
behaviour at small values of $y.$ Therefore, we conclude that in singlet
channel spin--independent part of ECT and spin--dependent one are
different yet in leading approximation, especially at small $y.$ This
effect takes place if created pair concentrated along the polarized
electron momentum direction and is absent (as we will see below) if pair
flies along unpolarized electron momentum direction.

\subsection{The contribution of the region $b)$}

\hspace{0.6cm}Considering double photon emission by longitudinally
polarized electron \cite{KM98} we saw that in the case when both photons
are pressed to the final (unpolarized) electron momentum direction the
spin--independent and spin--dependent parts of the ECT have quite the same
behaviour with the power accuracy. The analysis performed in this work
clarified that analogous situation takes place also for pair production.

We can explain our result by means of TD on Fig.2a. In the collinear region
$b)$ the (1,2) and (3,4) sets contribute. As before (in region $a)$) the
(1,2) set describes contribution into ECT due to nonsinglet channel, and
the (3,4) set now desribes the coresponding singlet channel contribution.
As we can see from Fig.2 TD's for (1,2) and (3,4) sets go one to another
at substitution
$$p_2 \ \leftrightarrow -p_+ \ .$$
This substitution does not affect polarized particle with 4--momentum
$p_1.$ That is why the contribution of both, spin--independent and
spin--dependent parts of the ECT, for (3,4) set can be obtained (on
differential level) from the corresponding contributions due to (1,2) set
by means the defined rule by analogy with Eqs.(8),(9). Because in
nonsinglet channel both parts of the ECT have the same behaviour, they
will be the same in singlet channel too. The interference of nonsinglet
channel and singlet one, which contributes with next--to--leading and
power accuracy, does not change this conclusion. At this
point the principal difference appears between the kinematics $a)$ and
$b)$ for pair production by polarized electron.

It is convenient in the region $b)$ to introduce the created electron and
positron energy fractions and angles relative to the final electron energy
and momentum direction, respectively
\begin{equation}\label{26, region b) variables}
y_{1,2} = \frac{x_{1,2}}{y_0} \ , \ \ y_0 =
\frac{\varepsilon_2}{\varepsilon} \ , \ \ \bar\theta_1 (\bar\theta_2) =
\widehat{\vec p_+\vec p_2} (\widehat{\vec p_-\vec p_2}) \ .
\end {equation}
The corresponding phase space of created pair in terms of these variables
reads
\begin{equation}\label{27, phase space in b)}
\frac{d^3p_+d^3p_-}{\varepsilon_+\varepsilon_-} = \pi^2m^4y_1y_2dy_1dy_2
d\bar z_1d\bar z_2\frac{d\phi}{2\pi}\ , \ \bar z_{1,2} =
y_0^2\frac{\varepsilon^2\bar{\theta_{1,2}}^2}{m^2} \ .
\end{equation}
We will define the collinear region $b)$ as a cone with the opening angle
$2\bar{\theta_0}$ along the scattered electron momentum direction,
therefore, the maximal value of $\bar z_{1,2}$ is $\bar z_0 =
y_0^2\frac{\varepsilon ^2\bar{\theta_{0}}^2}{m^2} \gg 1. $

According to the mention above we can write tensor $T_{\mu\nu}^{^{b)}}$
(see Eq.(5)) in the form
\begin{equation}\label{28, tensor in region b)}
T_{\mu\nu}^{^{b)}} = \frac{16}{m^4}I^{^{b)}}(Q_{\mu\nu} + i\lambda
E_{\mu\nu}) \ .
\end{equation}
In order to derive the quantity $I^{^{b)}}$ we have to at first find the
quantity $I^{^{a)}}$, which is the sum of all contributions accompanied by
the structure $Q_{\mu\nu}$ in right sides of Eqs. (7), (10) and (14)
and then use the rule \begin{equation}\label{29, rule for I(b)} I^{^{b)}}
= I^{^{a)}}(x_1\rightarrow -y_2, \ x_2\rightarrow -y_1, \ a_1\rightarrow
-b_2, \ a_2\rightarrow -b_1, \  a\rightarrow a, \ \Delta\rightarrow d )\ ,
\end{equation}
where
$$b_1 = \frac{1}{y_1}(1+y_1^2+y_1^2\bar z_1) \ , \ b_2 =
\frac{1}{y_2}(1+y_2^2+y_1^2\bar z_2) \ , \ d = a+b_1+b_2 \ . $$

The result of angular integration of the ECT in the region $b)$ can be
written as follows (see for comparison Eqs.(17),(18) and (19))
\begin{equation}\label{30, angular integration in b)}
\overline {L_{\mu\nu}^{b)}} = \frac{\alpha^2}{\pi^2}dy_1dy_2\ln\bar
z_0(Q_{\mu\nu} + i\lambda E_{\mu\nu})\overline{I^{b)}}
\end{equation}
$$\overline{I^{b)}} =
\frac{1+\eta^2}{(\eta-1)^4}\bigl[\frac{1}{2}(y_1^2+y_2^2)\ln\frac{\bar
z_0y_1^2y_2^2}{\eta^2} - (y_1-y_2)^2 + \frac{8\eta
y_1y_2}{1+\eta^2}\bigr]$$
$$+\frac{y_2^2+\eta^2}{(1+y_1)^4}\bigl[\frac{1}{2}(y_1^2+1)\ln\frac{\bar
z_0y_1^2y_2^2}{\eta^2} - (y_1-1)^2 + \frac{8\eta
y_1y_2}{y_2^2+\eta^2}\bigr]$$
$$+2\bigl[-\frac{y_1^2+\eta^2}{(1+y_1)(1-\eta)}\ln\frac{(\eta-1)(1+y_1)}
{\eta y_1} + \frac{y_1y_2-\eta}{(1+y_1)^2} + \frac{y_1-\eta
y_2}{(1-\eta)^2}\bigr] \ , $$
where $\eta = 1+y_1+y_2 .$

We can also integrate the right side of Eq.(30) over $y_2$ or $y_1$ at
fixed values of $\eta $: $y_1+y_2=\eta-1$ because in the region $b)$ the
4--momentum of the heavy photon depends on $\eta: \ q_a =p_2\eta -p_1 .$
The corresponding expression can be obtained using the symmetrical part of
$\widetilde L_{\mu\nu}^{a)}$ in Eq.(21) by the rule
\begin{equation}\label{31, spectra in b)}
\widetilde L_{\mu\nu}^{b)}(\eta\,,\bar z_0) = - sym\{\widetilde
L_{\mu\nu}^{a)}(y\,,z_0)\} \ ,
\end{equation}
and the result reads
$$\widetilde L_{\mu\nu}^{b)} = \frac{\alpha^2}{\pi^2}dy\ln\bar z_0\bigl\{
\ln\bar z_0\bigr[\frac{1+\eta^2}{3(\eta-1)} +
\frac{\eta-1}{6\eta}(4+7\eta+4\eta^2) - (1+\eta)\ln\eta\bigr]
-4(1+\eta)\ln\eta\ln(\eta-1)$$
$$- \frac{2}{\eta-1}\ln\eta - \frac{2(3\eta^2-1)}{\eta-1}L_{i2}(1-\eta)
- \frac{2}{3}\bigl[-4\eta^2-5\eta+1-\frac{4}{\eta(\eta-1)}\bigr]
\ln(\eta-1) - \frac{1}{3}\bigl(8\eta^2 +5\eta $$
\begin{equation}\label{32}
-7+\frac{13}{\eta-1}\bigr)\ln\eta + \frac{2}{3}\eta^2 - \frac{136}{9}\eta
+ \frac{107}{9} + \frac{4}{3\eta} - \frac{20}{9(\eta-1)}\bigr\}(Q_{\mu\nu}
+i\lambda E_{\mu\nu}) \ .
\end{equation}

At this point we want to pay attention that for a processes in which the
whole energy of the initial electron transforms into energy of the
electromagnetic jet along its momentum direction (the energy does not
transfer by heavy photon) variable $\eta$ equals to $1/y ,$ because in
this case $\varepsilon_2 = \varepsilon - \varepsilon_+ - \varepsilon_-.$
That allows to formulate the subsitution law (31) in terms of the same
variables: $y$ and $z_0.$ Such kind of law was used in calculation of QED
corrections to the small-angle Bhabha cross--section at LEP1 \cite{JETP95}.

\section{Semicollinear kinematics}

\hspace{0.6cm}Inside the investigated above collinear regions $a)$ and
$b)$ both, photon and fermion PD's, can be small as
compared with the heavy photon mass $q^2.$ In general case the smallness
of every PD gives the large logarithm in ECT after
corresponding angular integration.  Therefore, we have double--logarithmic
behaviour of the tensor $\widetilde L_{\mu\nu}$. Besides double
logarithmic terms the contribution of collinear regions contains also
single logarithm and constant relative to variable $z_0.$

The last kind of contribution can be arise also in kinematical regions
when all fermion PD's in underlying Feynman diagrams
have the same order as $q^2,$ and only photon ones leave small.
Such kinematics we call traditionally as semicollinear one.

It is easy to see that there are three semicollinear regions in the process
under consideration: $(\vec p_-\parallel \vec p_1)\ , \ (\vec p_-\parallel
\vec p_+)$ and $(\vec p_+\parallel \vec p_2).$ The corresponding
contributions into cross--sections for unpolarized initial electron were
studied in part for DIS \cite{NPM89} and small--angle Bhabha \cite{JETP95} processes.

Below we give the full analysis of these regions for longitudinally
polarized initial electron on the level of universal quantity -- ECT. We
use maximally the substitution laws based on topological equivalence of
essential TD's in every region to simplify the calculations. The final
result has a compact form, and we keep single logarithmic as well as
power contributions. In the frame of next--to--leading accuracy we
demonstrate the elimination of the angular auxiliary parameters
$\theta_0$ and $\bar\theta_0$ in the case when separation of collinear and
semicollinear regions has not physical sence.

\subsection{Contribution of the region $(\vec p_-\parallel \vec p_1)$}

\hspace{0.6cm}In the semicollinear region $(\vec p_-\parallel \vec p_1)$
the only (5,6) set of TD contributes. In this case the small photon
PD reads \begin{equation}\label{33, small propagator for 3.1}
q_1^2 = (p_- - p_1)^2 = - x_2m^2\bigl[\frac{(1-x_2)^2}{x_2^2} + z_2\bigr]
\ , \end{equation} and the phase space of created electron with
4--momentum $p_-$ is
\begin{equation}\label{34, phase space for 3.1}
\frac{d^3p_-}{\varepsilon_-} = \frac{1}{2}m^2x_2dx_2d\varphi dz_2 \ ,
\end{equation}
where $\varphi$ is the azimuth angle of the vector $\vec p_-$ in the
coordinate system with axis $OZ$ along the vector $\vec p_1.$

For investigation of ECT in this region it is convenient to introduce  the
"small" 4--vector
\begin{equation}\label{35, small vector in 3.1}
p = \frac{1}{x_2}p_- - p_1 \ ,
\end{equation}
which has only perependicular components in the choosen coordinate system.
When calculating the TD for (5,6) set we have to keep the terms of the
types
\begin{equation}\label{36, essential terms in 3.1}
\frac{1}{q_1^4}(m^2\, , q_1^2\, , (p,p))\ ,
\end{equation}
in order to achieve the adequate accuracy (including constant relative
$z_0$). The 4--vector $p$ can enter in ECT via scalar product and
tensor structures.

In accordance with (36) we can write the ECT in the considered region as
follows
\begin{equation}\label{37, general tensor in 3.1}
L_{\mu\nu}^{^{(3.1)}} =
\frac{\alpha^2}{2\pi^4}\frac{d^3p_+}{\varepsilon_+}x_2dx_2d\varphi dz_2
\frac{m^2}{q_1^4}\bigl[m^2L_{\mu\nu}^{^{m}} +
\frac{q_1^2}{(1-x_2)^2}L_{\mu\nu}^{^{(q_1)}} +
\frac{2x_2^2}{(1-x_2)^2}L_{\mu\nu}^{^{(p)}}\bigr] \ ,
\end{equation}
where the first term on the right side of Eq.(37) gives only constant
(next--next--to--leading) contribution relative $z_0$ (being
integrated over created electron angular phase space, see below) and reads
\begin{equation}\label{38,}
L_{\mu\nu}^{^{m}} = \frac{1}{ut_1}[(u+t_1)^2\tilde g_{\mu\nu} + 4q^2
\tilde p_{1\mu}\tilde p_{1\nu} - 2i\lambda E_{\mu\nu}^{^{m}}] \ .
\end{equation}
The second term inside the parenthesis in Eq.(37) leads to only logarithmic
(next--to--leading) contribution and is defined by the formula
\begin{equation}\label{39,}
L_{\mu\nu}^{^{(q_1)}} = \bigl[\frac{(1+x_2^2)(u^2+t_1^2)}{2ut_1} + 2x_2 +
\frac{q^2}{s_1}\bigr]\tilde g_{\mu\nu} - \frac{2q^2}{ut_1}[
(\tilde p_+\tilde p_2)_{\mu\nu} - (1+x_2^2)\tilde
p_{1\mu}\tilde p_{1\nu}] - \frac{i\lambda(1+x_2)}{ut_1}E_{\mu\nu}^{^{m}}
\ .
\end{equation}
At last, the third term on the right side of Eq.(37) leads to both,
logarithmic and constant, contributions: $$L_{\mu\nu}^{^{(p)}} =
-\frac{q^2}{2(1-x_2)^2}N^2g_{\mu\nu} - (KZ)_{\mu\nu} \ , \ \ q =
p_2+p_+-(1-x_2)p_1 \ , \ N = \chi_+ - \chi_2 \ , $$
\begin{equation}\label{40} K_{\mu} = \frac{N}{1-x_2}p_{2\mu} + \chi_2
p_{1\mu} + p_{\mu} \ , \ \ Z_{\mu} = - \frac{N}{1-x_2}p_{+\mu} + \chi_+
p_{1\mu} + p_{\mu} \ .
\end{equation}

To descibe the tensor $L_{\mu\nu}^{^{(3.1)}}$ we introduced the following
notation:
$$E_{\mu\nu}^{^{m}} =
(u-t_1)E_{\mu\nu}(q,p_2)+(s_1+(1-x_2)t_1)E_{\mu\nu}(q,p_1) \ , \
E_{\mu\nu}(a,b) = \epsilon_{\mu\nu\lambda\rho}a_{\lambda}b_{\rho} \ ,$$
$$u = -2p_1p_2\ , \ \ s_1 = 2p_2p_+ \ , \ \ t_1 = -2p_1p_+ \ , \ \
\chi_+ = \frac{2p_+p}{t_1}\ , \ \chi_2 = \frac{2p_2p}{u}\ , $$
\begin{equation}\label{41, notations in 3.1}
(ab)_{\mu\nu} = a_{\mu}b_{\nu} + a_{\nu}b_{\mu} \ , \ \tilde a_{\mu} =
a_{\mu}-\frac{aq}{q^2}q_{\mu} \ , \ \tilde g_{\mu\nu} = g_{\mu\nu} -
\frac{q_{\mu}q_{\nu}}{q^2} \ , \ \tilde aq = 0\ .
\end{equation}

Tensor $L_{\mu\nu}^{^{(p)}}$ satisfies condition:
$L_{\mu\nu}^{^{(p)}}q_{\nu} =0.$ Therefore, we could, in princple,
write it in terms of quantities with sign "tilde" as defined in the last
line of relations (41). But our strategy (as concerns this tensor) is at
first to integrate it over angular variables and then to write it by means
such kind quantities.

In the region $(\vec p_-\parallel\vec p_2)$ we can perform the
model--independent integration of the tensor
$L_{\mu\nu}^{^{(3.1)}}$ over angular
variables $z_2$ and $\varphi.$ The integration of the first two terms on
the right side of Eq.(37) is trivial and can be carried out by formulae
\begin{equation}\label{42, trivial integrals in 3.1}
\int\frac{m^4}{q_1^4}dz_2d\varphi = \frac{2\pi}{(1-x_2)^2} \ , \ \
\int\frac{m^2}{q_1^2}dz_2d\varphi = -
\frac{2\pi}{x_2}\ln\frac{z_0x_2^2}{(1-x_2)^2} \ .
\end{equation}

As concerns the integration of the third term in Eq.(37), it needs to use
the following relation
\begin{equation}\label{43, nontrivial integral in 3.1}
\int\frac{m^2}{q_1^4}dz_2d\varphi p_{\mu}p_{\nu} = -
\frac{\pi}{x_2^2}\bigl(\ln\frac{z_0x_2^2}{(1-x_2)^2} -
1\bigr)g_{\mu\nu}^{^{\perp}}\ ,
\end{equation}
where the perpendicular metric tensor $g_{\mu\nu}^{^{\perp}}$ has only
$xx$ and $yy$ components in the chosen coordinate system. It acts as
follows
\begin{equation}\label{44, action of g perp}
a_{\mu}g_{\mu\nu}^{^{\perp}} = a_{\nu}^{^{\perp}} , \ \
a_{\mu}b_{\nu}g_{\mu\nu}^{^{\perp}} = (ab)^{^{\perp}}\ , \ \
g_{\mu\nu}^{^{\perp}}p_{1\nu} = 0 \ .
\end{equation}

It is conveniet to write the Eq.(43) in the symbolic form as
\begin{equation}\label{45, symb. form}
p_{\mu}p_{\nu} \rightarrow - g_{\mu\nu}^{^{\perp}} \ .
\end{equation}
By using Eq.(43) in its symbolic form we can write
\begin{equation}\label{46, integrals}
\chi_+N \rightarrow \frac{2}{ut_1}(s_1+y_0t_1-x_1u) \ , \
\chi_2N \rightarrow \frac{2}{ut_1}(-s_1+y_0t_1-x_1u) \ ,
\end{equation}
$$Np_{\mu} \rightarrow \frac{2}{ut_1}(t_1p_{2\mu} - up_{+\mu} +(x_1u -
y_0t_1)p_{1\mu}) \ , $$
$$p_{\mu}p_{\nu} + \chi_2\chi_+p_{1\mu}p_{1\nu} +
(\chi_2+\chi_+)(p_1p)_{\mu\nu} \rightarrow - g_{\mu\nu} - \frac{2s_1}{ut_1}
p_{1\mu}p_{1\nu} - \frac{(up_+ +t_1p_2,p_1)_{\mu\nu}}{ut_1} \ . $$
When writing the last relation in (46) we used the following
representation of the metrical tensor
\begin{equation}\label{47 metrical tensor}
g_{\mu\nu} = g_{\mu\nu}^{^{\perp}} +
\frac{1}{\varepsilon^2}p_{1\mu}p_{1\nu} + \frac{1}{\varepsilon}(g_{\mu
z}p_{1\nu} + g_{\nu z}p_{1\mu}) \ .
\end{equation}

Looking at relations (46), one can see that the result of angular
integration of the tensor $L_{\mu\nu}^{^{(p)}}$ in considered region can
be written in covariant form. Because of gauge invariance, we can introduce
the quantities with "tilde" and use the equation $$\tilde p_+ + \tilde p_2
-(1-x_2)\tilde p_1 = 0$$ at this point. The result has a very simple form
\begin{equation}\label{48, }
L_{\mu\nu}^{^{(p)}} \rightarrow -2\bigl(\frac{q^2s_1}{(1-x_2)^2ut_1}
-1\bigr)\tilde g_{\mu\nu} + \frac{4q^2}{(1-x_2)^2ut_1}(\tilde p_2\tilde
p_+)_{\mu\nu} \ , \ q^2 = s_1 +(1-x_2)(u+t_1) \ .
\end{equation}

The Eqs.(42) and (43) indicate that the expression for contribution of the
semicollinear region $(\vec p_-\parallel\vec p_2)$ into ECT can be
written as follows
\begin{equation}\label{49, full ECT in 3.1}
\int
L_{\mu\nu}^{^{(3.1)}} =
\frac{\alpha^2}{\pi^3}\frac{d^3p_+}{\varepsilon_+}\frac{dx_2}{(1-x_2)^2ut_1}
\bigl[x_2a_{\mu\nu} + b_{\mu\nu}\ln\frac{z_0x_2^2}{(1-x_2)^2}\bigr] \ ,
\end{equation}
where
\begin{equation}\label{50}
a_{\mu\nu} = c\tilde g_{\mu\nu} + \frac{4q^2}{(1-x_2)^2}(
\tilde p_{2\mu}\tilde p_{2\nu} + \tilde p_{+\mu}\tilde p_{+\nu}) -
2i\lambda E_{\mu\nu}^{^{m}} \ , \ c = u^2+t_1^2 +
\frac{2q^2s_1}{(1-x_2)^2}\ ,
\end{equation}
\begin{equation}\label{51}
b_{\mu\nu} = (1+x_2^2)\bigl[-\frac{c}{2}\tilde
g_{\mu\nu}-\frac{2q^2}{(1-x_2)^2}(\tilde p_{2\mu}\tilde p_{2\nu} +
\tilde p_{+\mu}\tilde p_{+\nu})\bigr] +i\lambda(1+x_2)E_{\mu\nu}^{^{m}}\ .
\end{equation}

For situations in which the created electron--positron pair is not
observed we have to sum the contributions due to collinear and
semicollinear regions. In such cases parameter $\theta_0,$ which separates
these regions has not deep physical sence and must disappear in the final
expression for any observed physical quantity. In the frame of
next--to--leading accuracy this fact leads to cancellation of all terms
proportional to $\ln\theta_0^2\ln\frac{\varepsilon^2}{m^2}.$
%\end{document}
Let us show that such cancellation takes place for contribution of the
(5,6) set of TD. In order to extract the corresponding term when
integrating the right side of Eq.(49) over created positron angular phase
space
\begin{equation}\label{52, created positron phase space in 3.1}
\frac{d^3p_+}{\varepsilon_+} = \varepsilon^2x_1dx_1d\varphi_+dc_1 \ ,
\end{equation}
(where $c_1 = \cos{\theta_1}$ and $\varphi_+$ is the positron azimuth
angle) it is convenient to write tensor $b_{\mu\nu}$ on the right side of
Eq.(49) as follows
\begin{equation}\label{53, tensor b in 3.1}
b_{\mu\nu}(c_1) = [b_{\mu\nu}(c_1) - b_{\mu\nu}(1)] + b_{\mu\nu}(1) \ .
\end{equation}
The upper limit of integration over $c_1$ equals to
$\cos{\theta_0} .$
%Because we neglect with $\theta_0^2$ as compared

%with unity we must integrate
Only the pole--like term proportional to $b_{\mu\nu}(1)/t_1$ on the right
side of Eq.(49) gives contribution, we want to extract. By using
\begin{equation}\label{54}
q^2 = yu\ , \ \ s_1 = -x_1u\ , \ \ y\tilde p_1 = \tilde p_2
\end{equation}
for $b_{\mu\nu}(1)$ we obtain the interesting contribution in the form
\begin{equation}\label{55}
-\frac{\alpha^2}{\pi^2}dx_1dx_2\ln{\theta_0^2}\ln\frac{\varepsilon^2}{m^2}
\bigl[\frac{(1+x_2^2)(y^2+x_1^2)}{(1-x_2)^4}Q_{\mu\nu} + i\lambda\frac
{(y-x_1)(1+x_2)}{(1-x_2)^2}E_{\mu\nu}\bigr] \ .
\end{equation}

It is enough to look at Eqs.(17),(19) and (20) to see that all terms which
contain $\ln\theta_0^2\ln\frac{\varepsilon^2}{m^2}$ vanish when we sum
collinear and semicollinear contributions due to (5,6) set of TD. Usually
the contribution of semicollinear regions being added to collinear one
restores argument of the leading logarithm in a such a way that $\ln^2z_0$
transforms into $\ln^2(-u/m^2)$ (see, for example, \cite{JETP95}). This
restoration is connected also with contribution of the lower limit
integration of the pole--like term that depends on concrete physical
applications. Below we will not consentrate on this point more.

\subsection{The contribution of the region $(\vec p_-\parallel\vec p_+)$}

\hspace{0.6cm}In the semicollinear region $(\vec p_-\parallel\vec p_+)$
only (1,2) set of TD contributes. The small photon PD in this
region has virtuality \begin{equation}\label{56, small propagator in
3.2} q_2^2 = (p_+ + p_-)^2 = m^2x_1x_2\bigl[\frac{(1-y)^2}{x_1^2x_2^2} +
z\bigr] \ , \ \ z = \frac{\varepsilon^2\theta_{12}^2}{m^2} \ ,
\end{equation} where $\theta_{1,2} = \widehat{\vec p_+\vec p_-}$ is the
angle between the created electron and positron momentum directions. The
phase space of created electron can be parametrized as follows
\begin{equation}
\frac{d^3p_-}{\varepsilon_-} = \frac{m^2}{2}x_2dx_2d\varphi_-dz \ ,
\end{equation}
where now $\varphi_-$ is the azimuth angle of the vector $\vec p_-$ in the
system with axis $OZ$ along the direction of vector $\vec p_+. $

If we will introduce the "small" 4--vector
\begin{equation}\label{58, small vector in 3.2}
h = \frac{x_1}{x_2}p_- - p_+ \ ,
\end{equation}
we can write the ECT in the considered region by full analogy with
Eq.(37):  \begin{equation}\label{59, general tensor in 3.1}
L_{\mu\nu}^{^{(3.2)}} =
\frac{\alpha^2}{2\pi^4}\frac{d^3p_+}{\varepsilon_+}x_2dx_2d\varphi_- dz
\frac{m^2}{q_2^4}
\bigl[m^2\bar L_{\mu\nu}^{^{m}} +
\frac{q_2^2x_1^2}{(1-y)^2}L_{\mu\nu}^{^{(q_2)}} +
\frac{2x_2^2}{(1-y)^2}L_{\mu\nu}^{^{(h)}}\bigr] \ .
\end{equation}

As it follows from Fig.2a the (1,2) set of TD can be obtained from (5,6)
one by the interchange 4--momenta $p_1$ and $-p_+.$ Such operation
changes the conditions of polarized particle. That is why by means the
corresponding substitution law (see Eq.(9))
\begin{equation}\label{60, substitution law in 3.2}
\hat P = [-p_+ \leftrightarrow p_1\ , \ (x_1\ , x_2\ , y_0 \ , z_2 \ , p)
\rightarrow \bigl(\frac{1}{x_1}\ , \frac{-x_2}{x_1}\ , \frac{-y_0}{x_1}\ ,
x_1^2z \ , - h\bigr)]\ ,
\end{equation}
$$\hat P (u\ ,s_1\ ,t_1) \rightarrow (s_1\ ,u\ ,t_1)$$
we can derive from Eq.(37) only symmetrical spin--independent part of the
ECT for (1,2) set
\begin{equation}\label{61, symmetrical 3.2}
sym\bigl[m^2\bar L_{\mu\nu}^{^{m}} +
\frac{q_2^2x_1^2}{(1-y)^2}L_{\mu\nu}^{^{(q_2)}} +
\frac{2x_2^2}{(1-y)^2}L_{\mu\nu}^{^{(h)}}\bigr] =
\hat P sym\bigl[m^2L_{\mu\nu}^{^{m}} +
\end{equation}
$$
\frac{q_1^2}{(1-x_2)^2}L_{\mu\nu}^{^{(q_1)}} +
\frac{2x_2^2}{(1-x_2)^2}L_{\mu\nu}^{^{(p)}}\bigr] \ . $$
The antisymmetrical spin--dependent part of ECT requires of independent
calculations. The result can be written as follows:
\begin{equation}\label{62,}
\bar L_{\mu\nu}^{^{m}} = \frac{1}{s_1t_1}[(s_1+t_1)^2\tilde g_{\mu\nu} +
4Q^2 \tilde p_{+\mu}\tilde p_{+\nu} - 2i\lambda(t_1-s_1)
E_{\mu\nu}(Q,p_+)] \ ,
\end{equation}
$$Q = p_2-p_1+\frac{1-y}{x_1}p_+ \ , \ \ \tilde a_{\mu} = a_{\mu} -
\frac{aQ}{Q^2}Q_{\mu} \ ,$$

$$L_{\mu\nu}^{^{(q_2)}} = \frac{1}{x_1^2s_1t_1}\bigl\{
\bigl[\frac{(x_1^2+x_2^2)(s_1^2+t_1^2)}{2} - 2x_1x_2s_1t_1 +
x_1^2Q^2u\bigr]\tilde g_{\mu\nu} + 2Q^2[
x_1^2(\tilde p_1\tilde p_2)_{\mu\nu} + (x_1^2+x_2^2)\tilde
p_{+\mu}\tilde p_{+\nu}]$$
\begin{equation}\label{63}
- i\lambda x_1[(x_1u+(1-y)t_1)E_{\mu\nu}(Q,p_1) + (x_1u +
s_1(1-y))E_{\mu\nu}(Q,p_2) + 2x_2(s_1-t_1)E_{\mu\nu}(Q,p_+)]\bigr\} \ ,
\end{equation}
$$
L_{\mu\nu}^{^{(h)}} = -\frac{x_1^2Q^2}{2(1-y)^2}N_h^2g_{\mu\nu} -
(K^hZ^h)_{\mu\nu} + i\lambda\frac{N_hx_1^2}{(1-y)^2}\bigl[\chi_1E_{\mu\nu}
(Q,p_2)
$$
\begin{equation}\label{64,}
+\chi_2^hE_{\mu\nu} -\frac{1-y}{x_1}E_{\mu\nu}(Q,h)\bigr] \ , \ \ N^h =
\chi_1 + \chi_2^h \ , \ \ \chi_1=\frac{2p_1h}{t_1}\ , \ \ \chi_2^h =
\frac{2p_2h}{s_1} \ ,
\end{equation}
$$K_{\mu}^h = \frac{x_1N^h}{1-y}p_{2\mu} +
\chi_2^h p_{+\mu} - h_{\mu} \ , \ \ Z_{\mu}^h =
\frac{x_1N^h}{1-y}p_{1\mu} - \chi_1 p_{+\mu} - h_{\mu} \ .  $$

The tensor $L_{\mu\nu}^{^{(h)}}$ satisfies the
condition $L_{\mu\nu}^{^{(h)}}Q_{\nu} = 0 $ because of gauge invariance.
Of course, we can exclude the sructure $E_{\mu\nu}(Q,p_+)$ using the
definition of 4--vector $Q$ given in (62). But the written expressions on
the right sides of Eqs.(62) and (63) are more compact in our opinion.

The angular integration of the tensor $L_{\mu\nu}^{^{(3.2)}}$ can be
carried out by the help of relations
\begin{equation}\label{65, integrals in 3.2}
\int\frac{m^4}{q_2^4}dzd\varphi_- = \frac{2\pi}{(1-y)^2} \ , \ \
\int\frac{m^2}{q_2^2}dzd\varphi_- =
\frac{2\pi}{x_1x_2}\ln\frac{z_ax_1^2x_2^2}{(1-y)^2} \ ,
\end{equation}
$$\int\frac{m^2}{q_2^4}dzd\varphi_- h_{\mu}h_{\nu} = -
\frac{\pi}{x_2^2}\bigl(\ln\frac{z_ax_1^2x_2^2}{(1-y)^2} -
1\bigr)g_{\mu\nu}^{^{\perp}}\ ,$$
where $ z_a = \varepsilon^2\theta_a^2/m^2$, and the perpendicular metric
tensor $g_{\mu\nu}^{^{\perp}}$ has only $xx$ and $yy$ components in
the chosen coordinate system (the axis $OZ$ along vector $\vec p_+$). We
introduced here parameter $\theta_a \ll 1$ which defines the
semicollinear region $(\vec p_-\parallel\vec p_+)$ in such a way that
$\theta_{1,2} \leq \theta_a.$ Note, that Eqs.(65) can be derived from the
corresponding Eqs.(42) and (43) by applicaton of operation $(1/x_1^2)\hat
P.$

Using the symbolic form of the last relation in (65) we can write
$$\chi_2^hN_h \rightarrow \frac{2}{x_1s_1t_1}(x_1u + s_1 + y_0t_1) \ , \ \
\chi_1N_h \rightarrow \frac{2}{x_1s_1t_1}(x_1u - s_1 - y_0t_1) \ ,$$
$$N_hh_{\mu} \rightarrow -\frac{2}{s_1t_1}\bigl(t_1p_{2\mu}+s_1p_{1\mu}
-\frac{s_1+y_0t_1}{x_1}p_{+\mu}\bigr)\ ,$$
\begin{equation}\label{66, hh int in 3.2}
h_{\mu}h_{\nu} - \chi_2^h\chi_1p_{+\mu}p_{+\nu} +
(\chi_1-\chi_2^h)(p_+h)_{\mu\nu} \rightarrow - g_{\mu\nu} -
\frac{2u}{s_1t_1} p_{+\mu}p_{+\nu} - \frac{(s_1p_1
-t_1p_2,p_+)_{\mu\nu}}{s_1t_1} \ .
\end{equation}

It is easy to see that Eqs.(66) follow from Eqs.(46) after application of
operator $\hat P$ to the last ones. The metric tensor in (66) is defined
by analogy with (47):
\begin{equation}\label{67 metrical tensor in 3.2}
g_{\mu\nu} = g_{\mu\nu}^{^{\perp}} +
\frac{1}{\varepsilon_+^2}p_{+\mu}p_{+\nu} + \frac{1}{\varepsilon_+}(g_{\mu
z}p_{+\nu} + g_{\nu z}p_{+\mu}) \ .
\end{equation}
By means of Eqs.(66) we derive
\begin{equation}\label{68, tensor hh in 3.2}
L_{\mu\nu}^{^{(h)}} \rightarrow -2\bigl(\frac{x_1^2Q^2u}{(1-y)^2s_1t_1}
-1\bigr)\tilde g_{\mu\nu} - \frac{4x_1^2Q^2}{(1-y)^2s_1t_1}(\tilde
p_1\tilde p_2)_{\mu\nu}
\end{equation}
$$+
2i\lambda\frac{x_1}{(1-y)^2s_1t_1}[(x_1u+(1-y)t_1)E_{\mu\nu}(Q,p_2)
+ (x_1u +(1-y)s_1)E_{\mu\nu}(Q,p_1)] \ . $$

The general expression for the contribution of the semicollinear region
$(\vec p_-\parallel\vec p_+)$ into ECT can be written as follows
\begin{equation}\label{69, full ECT in 3.1}
\int L_{\mu\nu}^{^{(3.2)}} =
\frac{\alpha^2}{\pi^3}\frac{d^3p_+}{\varepsilon_+}\frac{dx_2}{(1-y)^2s_1t_1}
\bigl[x_2a_{\mu\nu}^{^{(1)}} +
b_{\mu\nu}^{^{(1)}}\ln\frac{z_ax_1^2x_2^2}{(1-y)^2}\bigr] \ ,
\end{equation}
\begin{equation}\label{70}
a_{\mu\nu}^{^{(1)}} = c^{^{(1)}}\tilde g_{\mu\nu} +
\frac{4x_1^2Q^2}{(1-y)^2}( \tilde p_{2\mu}\tilde p_{2\nu} + \tilde
p_{1\mu}\tilde p_{1\nu}) - 2i\lambda\frac{x_1}{(1-y)^2}\bigl[(ux_1 +
(1-y)s_1 E_{\mu\nu}(Q,p_2)
\end{equation}
$$+ (ux_1 + (1-y)t_1)E_{\mu\nu}(Q,p_1)\bigr] \ , $$
\begin{equation}\label{71}
b_{\mu\nu}^{^{(1)}} = \frac{
(x_1^2+x_2^2)}{x_1}\bigl\{\frac{c^{^{(1)}}}{2}\tilde
g_{\mu\nu} + \frac{2x_1^2Q^2}{(1-y)^2}( \tilde p_{2\mu}\tilde
p_{2\nu} + \tilde p_{1\mu}\tilde p_{1\nu}) -
i\lambda\frac{x_1}{(1-y)^2}\bigl[(ux_1 + (1-y)t_1)E_{\mu\nu}(Q,p_1)
\end{equation}
$$+ (ux_1 + (1-y)s_1)E_{\mu\nu}(Q,p_2)\bigr]\bigr\} \ , \ c^{^{(1)}} =
s_1^2+t_1^2 + \frac{2x_1^2}{(1-y)^2}uQ^2 \ . $$

The spin--independent part of ECT on differential level given by 
Eqs.(61)--(64) coincides with the corresponding results of Ref.[8,9].
But the integration of ECT in [8] has been performed with mistakes. Here 
we conclude that formula (17) of Ref.[8] (which is analog of our formula 
(69) for spin--independent part of ECT) is incorrect.

The ECT in the region $(\vec p_-\parallel\vec p_+)$ has a pole--like 
behaviour at both, small $t_1$ and small $s_1.$ One can verify that terms 
proportional to $\ln\theta_0^2\ln\frac{\varepsilon^2}{m^2}$ cancel in the 
sum of contributions of the semilollinear region $(\vec p_-\parallel\vec 
p_+)$ at small $t_1$ and collinear region $a),$ while terms proportional 
to $\ln\bar\theta_0^2\ln\frac{\varepsilon^2}{m^2}$ cancel due to 
contributions of region $(\vec p_-\parallel\vec p_+)$ at small $s_1$ and
collinear region $b).$

In the limiting case $|t_1| \ll |u|\ , s_1\ , |Q^2|$ we can extract the
coresponding terms using Eqs.(69) and (71) as well as relations (54) in the
same way as it was done above in Section 3.1, and the result reads
\begin{equation}\label{72}
-\frac{\alpha^2}{\pi^2}dx_1dx_2\ln{\theta_0^2}\ln\frac{\varepsilon^2}{m^2}
\frac{(x_1^2+x_2^2)(y^2+1)}{(1-y)^4}[Q_{\mu\nu} + i\lambda E_{\mu\nu}] \ .
\end{equation}
Looking at Eqs.(17) and (18) we see that the corresponding contribution
due to (1,2) set of TD in the collinear region $a)$ has just opposite sign
as compared with expression (72).

In another limiting case $s_1 \ll |u|\ , |t_1|\ , |q^2|$ we have to
use relations
\begin{equation}\label{73}
t_1 = y_1u \ , \ Q^2 = \eta u \ , \ \tilde p_+ = y_1\tilde p_2 \ , \
s_1 = 2\varepsilon^2y_1(1-\bar c_1)\ , \ \bar c_1 = \cos{\bar\theta_1}
\end{equation}
to compute the quantity $b_{\mu\nu}^{^{(1)}}$ at $c_1 =1$ and derive
\begin{equation}\label{74}
-\frac{\alpha^2}{\pi^2}dy_1dy_2\ln{\bar\theta_0^2}\ln\frac{\varepsilon^2}{m^2}
\frac{(y_1^2+y_2^2)(\eta^2+1)}{(1-\eta)^4}[Q_{\mu\nu} + i\lambda
E_{\mu\nu}] \ .
\end{equation}
The expression (74) cancels the contribution contaiting
$\ln{\bar\theta_0^2}\ln\frac{\varepsilon^2}{m^2}$ due to (1,2) set of TD
in collinear region $b)$ as it follows from Eqs.(30),(31).

\subsection{Contribution of the region $(\vec p_+\parallel\vec p_2)$}

\hspace{0.6cm}In the semicollinear region $(\vec p_+\parallel\vec p_2)$
only (3,4) set of TD contributes. As one can see from Fig.2a the TD of
that set may be obtained from (1,2) one by interchange
$p_2\leftrightarrow p_-.$ The such kind of substitution does not affect
any condition for polarized particle with 4--momentum $p_1.$ Thetefore,
both, spin--independent part of the corresponding tensor
$L_{\mu\nu}^{^{(3.3)}}$ and spin--dependent one, can be derived by definite
subsitution law using tensor $L_{\mu\nu}^{^{(3.2)}}.$ If we will write
\begin{equation}\label{75} L_{\mu\nu}^{^{(3.3)}} =
\frac{\alpha^2}{2\pi^4}\frac{d^3p_-}{\varepsilon_-}
y_1dy_1\frac{m^2}{q_3^4}d\bar z_1d\varphi_+ L_{\mu\nu} \ , \ q_3^2 =
(p_2+p_+)^2
\end{equation}
then tensor $L_{\mu\nu}$ on the right side of Eq.(75) can be written by
means of the right hand side of Eq.(59) in the form
\begin{equation}\label{76}
L_{\mu\nu} = \widehat{O} \bigl[m^2\bar L_{\mu\nu}^{^{(m)}} +
\frac{q_2^2x_1^2}{(1-y)^2}l_{\mu\nu}^{^{(q_2)}} + \frac{2x_2^2}{(1-y)^2}
L_{\mu\nu}^{^{(h)}}\bigr] \ ,
\end{equation}
where operator $\widehat{O}$ is defined as follows
\begin{equation}\label{77}
\widehat{O} = [p_2\leftrightarrow p_-\ , \ (p_+\ , x_2\ , x_1\ , h)
\rightarrow (y_1p_2\ , y_0\ , x_1\ , -y_1f)]\ , \ f=\frac{1}{y_1}p_+ - p_2
\ .  \end{equation} (For the notation used here see Section 2.2) The
action of the operator $\widehat O$ on the used invariants reads
\begin{equation}\label{78}
\widehat{O} (t_1\ , s_1\ , u\ , q_2\ , Q ) = (y_1u\ , y_1s_2\ , t_2\ , q_3\
\bar{Q}) \ ,
\end{equation}
$$s_2 = 2p_2p_-\ , \ \ t_2 = - 2p_1p_-\ , \ \ \bar{Q} = p_2(1+y_1) +p_-
-p_1\ . $$
We omitt all intermediate calculations and give only the final result for
ECT in the region $(\vec p_+\parallel\vec p_2)$
\begin{equation}\label{79}
\int L_{\mu\nu}^{^{(3.3)}} =
\frac{\alpha^2}{\pi^3}\frac{d^3p_-}{\varepsilon_-}\frac{dy_1}{(1+y_1)^2us_2}
\bigl[y_1a_{\mu\nu}^{^{(2)}} +
b_{\mu\nu}^{^{(2)}}\ln\frac{\bar z_0y_1^2}{(1+y_1)^2}\bigr] \ ,
\end{equation}
\begin{equation}\label{80}
a_{\mu\nu}^{^{(2)}} = c^{^{(2)}}\tilde g_{\mu\nu} +
\frac{4\bar{Q^2}}{(1+y_1)^2}( \tilde p_{-\mu}\tilde p_{-\nu} + \tilde
p_{1\mu}\tilde p_{1\nu}) - 2i\lambda\frac{1}{(1+y_1)^2}\bigl[(t_2 +
(1+y_1)s_2)E_{\mu\nu}(\bar{Q},p_-)
\end{equation}
$$+ (t_2 + (1+y_1)u)E_{\mu\nu}(\bar{Q},p_1)\bigr] \ , $$
\begin{equation}\label{81}
b_{\mu\nu}^{^{(2)}} =
(1+y_1^2)\bigl\{\frac{c^{^{(2)}}}{2}\tilde
g_{\mu\nu} + \frac{2\bar{Q^2}}{(1+y_1)^2}( \tilde p_{-\mu}\tilde
p_{-\nu} + \tilde p_{1\mu}\tilde p_{1\nu}) -
i\lambda\frac{1}{(1+y_1)^2}\bigl[(t_2 + (1+y_1)u)E_{\mu\nu}(\bar{Q},p_1)
\end{equation}
$$+ (t_2 + (1+y_1)s_2)E_{\mu\nu}(\bar{Q},p_-)\bigr]\bigr\} \ , \ c^{^{(2)}}
= s_2^2+u^2 + \frac{2t_2\bar Q^2}{(1+y_1)^2} \ . $$

In the limiting case $s_2\ll|t_2|\ , |u|\ , |\bar{Q^2}|$ we have to use
relations
\begin{equation}\label{82}
t_2 = y_2u\ , \bar{Q^2} = \eta u\ ,\ s_2 = 2\varepsilon_2y_2(1-\bar c_2)\ ,
\ \bar c_2 = \cos{\bar\theta_2} \
\end{equation}
to compute the quantity $b_{\mu\nu}^{^{(2)}}$ at $\bar c_2 =1$
and extract (by analogy with (55)) term proportional to $\ln\bar\theta_0^2
\ln\frac{\varepsilon^2}{m^2}.$ It reads
\begin{equation}\label{83}
-\frac{\alpha^2}{\pi^2}dy_1dy_2\ln{\bar\theta_0^2}\ln\frac{\varepsilon^2}{m^2}
\frac{(1+y_1^2)(\eta^2+y_2^2)}{(1+y_1)^4}[Q_{\mu\nu} + i\lambda
E_{\mu\nu}] \ .
\end{equation}
Expression (83) cancels exactly the corresponding contribution of the
collinear region $b)$ due to (3,4) set of TD (see Eq.(30)).

\section{Conclusion}

\hspace{0.6cm}In this paper we calculated the ECT for the process of
electron--positron pair production at the scattering of longitudinally
polarized electron on heavy photon. The work was stimulated by recent
polarized experiments on deep inelastic scattering [1,2], but the obtained
result can be used to compute the second order radiative correction due to
hard pair production for a wide class of the scattering and annihilation
processes. The contribution of collinear and semicollinear kinematical
regions are studied, and that allows to find the corresponding correction
with the next--to--leading accuracy. The cancellation of the angular
auxiliary parameters $\theta_0$ and $\bar{\theta_0}$ in the case of
unobserved created pair indicates that our results for collinear regions
are in accordance with semicollinear ones. We give together
spin--independent and spin--dependent parts of ECT to make polarization
effects more transparent on the level of theoretical formulae.

We want to pay attention that in contrast with the purely photonic
corrections the leading correction, connected with hard pair production,
has different form for spin--depedent and spin--independent parts of the
ECT because of contribution in singlet channel. This fact indicates that
for asymmetry--like quantities the full second--order correction will
dominate just due to pair production via singlet channel because the whole
leading nonsinglet channel contribution cancels in this case \cite{AAK98}.
The last includes all photonic corrections and nonsinglet part of
corrections connected with pair production. \\

{\large{\bf{Acknowledgements}}}  \\

\hspace{0.6cm}The authors thanks G.I. Gakh, A.B. Arbuzov and L.
Trentadue for a discussion. This work supported in part by INTAS Grant
93--1867 ext and Ukrainian DFFD grant 2.4/379. \\

\section*{Appendix}
\setcounter{equation}{0}
\renewcommand{\theequation}{A.\arabic{equation}}

\hspace{0.6cm}In this Appendix we want outline the starting--points for
calculation of tensor $T_{\mu\nu}$ given on the right side of Eq.(5). In
the considered case module of heavy photon viruality $|q^2|$ and invariant
$u=-2p_1p_2,$ that defines the registration conditions of the scattered
electron, are much more as compared with $m^2.$ Therefore, only diagrams
on Fig.1 contribute in collinear and semicollinear knematics, and tensor
$T_{\mu\nu}$ in general can be written as follows
\begin{equation}\label{A1}
T_{\mu\nu} = (M^{^{(12)}} - M^{^{(34)}} - M^{^{(56)}})_{\mu}
(M^{^{(12)}} - M^{^{(34)}} - M^{^{(56)}})_{\nu}^{^{+}} \ ,
\end{equation}
$$M^{^{(12)}} = \bar{u}(p_2)Q_{\mu\lambda}^{^{(12)}}u(p_1)\frac{1}{q_2^2}
\bar{u}(p_-)\gamma_{\lambda}v(p_+) \ , \ \ M_{\mu}^{^{(34)}} = \widehat{O}
M^{^{(12)}}\ , \ \ M_{\mu}^{^{(56)}} = \widehat{P}M^{^{(12)}}\ , $$
\begin{equation}\label{A2}
Q_{\mu\nu}^{^{(12)}} = \gamma_{\mu}\frac{\hat
{p_1}-\hat{q_2}+m}{q_2^2-2p_1q_2}\gamma_{\lambda} +
\gamma_{\lambda}\frac{\hat
{p_2}+\hat{q_2}+m}{q_2^2+2p_2q_2}\gamma_{\mu}\ ,
\end{equation}
where we used the notation of Section 3.

In the colliear region $a)$ the (3,4) set does not contribute, and we have
\begin{equation}\label{A3}
T_{\mu\nu}^{^{a)}} = M_{\mu}^{^{(12)}}M_{\nu}^{^{(12)+}} +
M_{\mu}^{^{(56)}}M_{\nu}^{^{(56)+}}
-(M_{\mu}^{^{(12)}}M_{\nu}^{^{(56)+}}+M_{\mu}^{^{(56)}}M_{\nu}^{^{(12)+}})\ .
\end{equation}
The first term on the right side of Eq.(A.3) corresponds to contribution
of (1,2) set of TD on Fig.2a that describes the nonsinglet channel only. It
reads
\begin{equation}\label{A.4}
M_{\mu}^{^{(12)}}M_{\nu}^{^{(12)+}} =
\frac{1}{q_2^4}[-2q_2^2g_{\lambda\rho} + 4(p_+p_-)_{\lambda\rho}]Tr(\hat
p_2+m)Q_{\mu\lambda}^{^{(12)}}(\hat
p_1+m)\bigl(1+\lambda\gamma_5\bigl(1+\frac{m\hat k}{p_1k}\bigr)\bigr)
Q_{\nu\rho}^{^{(12)+}}\ ,
\end{equation}
where we define the initial electron polarization 4--vector $a_{\mu}$ in
the form \cite{KM98,BKL83}
$$a_{\mu} = \frac{\lambda}{m}\bigl(p_1-\frac{m^2k}{p_1k}\bigr)\ , \ \
k= (\varepsilon, \ -\vec p_1)$$
and used the relation
$$(\hat p_1+m)\bigl[1-\frac{\lambda}{m}\gamma_5\bigl(\hat
p_1-\frac{m^2\hat k}{p_1k}\bigr)\bigr] = (\hat
p_1+m)\bigl[1+\lambda\gamma_5\bigl(1+\frac{m\hat k}{p_1k}\bigr)\bigr]\ .
$$
Note that vector $k$ vanishes in final results because in the frame of
choosen accuracy it contributes via scalar production $(kp_1)$ in the same
way as for double--photon emission \cite{KM98}. The multiplier inside
the right brackets on the right side of Eq.(A.4) describes the upper block
of the corresponding TD Fig.2a while the trace -- lower one. The part of
trace, that contains the doubled initial electron helicity $\lambda,$ is
symmetric relative indecies $(\lambda,\rho)$ and antisymmetric relative
$(\mu,\nu)$ ones.

The second term on the right side of Eq.(A.3) is responsibe for the
singlet channel contribution. It connected with (5,6) set of TD and can be
written as follows \begin{equation}\label{A.5}
M_{\mu}^{^{(56)}}M_{\nu}^{^{(56)+}} =
\frac{1}{q_1^4}\bigl[2q_1^2g_{\lambda\rho}+4(p_1p_-)_{\lambda\rho}+4i\lambda
\bigl(E_{\lambda\rho}(p_1,p_-) +
\frac{m^2}{p_1k}E_{\lambda\rho}(p_--p_1,k)\bigr)\bigr]
\end{equation}
$$\times Tr(\hat p_2+m)Q_{\mu\lambda}^{^{(56)}}(\hat p_+ -
m)Q_{\nu\rho}^{^{(56)+}} \ .$$

Now the polarized electron belongs to upper block of TD, and we see that
the corresponding expression includes both symmetric and antisymmetric
parts. To derive the spin--dependent part of (5,6) set we must compute in
this case the antisymmetrical relative both pairs of indecies
$(\lambda,\rho)$ and $(\mu,\nu)$ part of the trace on the right side of
Eq.(A.5). Only spin--independent parts on the right side of Eqs.(A.4) and
(A.5) transform one to another by action of operator $\widehat P$ at which
$p_1\leftrightarrow-p_+.$ As concerns spin--dependent ones, they must be
calculated independently.

The third term (in the paranthesis) on the right side of Eq.(A.3) describes
the interference of singlet and nonsinglet channels. It corresponds to
another class of TD (Fig.2,b). If we represent it as a sum of symmetrical
and antisymmetrical parts, then
$$sym\{M_{\mu}^{^{(12)}}M_{\nu}^{^{(56)+}}+
M_{\mu}^{^{(56)}}M_{\nu}^{^{(12)+}}\} = M_{\mu}^{^{(12)}}M_{\nu}^{^{(56)+}}
+(\mu\leftrightarrow\nu) \ , $$
$$asym\{M_{\mu}^{^{(12)}}M_{\nu}^{^{(56)+}}+
M_{\mu}^{^{(56)}}M_{\nu}^{^{(12)+}}\} = M_{\mu}^{^{(12)}}M_{\nu}^{^{(56)+}}
-(\mu\leftrightarrow\nu) \ , $$
where
\begin{equation}\label{A.6}
M_{\mu}^{^{(12)}}M_{\nu}^{^{(56)+}} = \frac{1}{q_1^2q_2^2}(\hat
p_2+m)Q_{\mu\lambda}^{^{(12)}}(p_1+m)\bigl[1+\lambda\gamma_5\bigl(1+
\frac{m\hat k}{p_1k}\bigr)\bigr]\gamma_{\rho}(\hat p_-+m)\gamma_{\lambda}
(\hat p_+-m)Q_{\nu\rho}^{^{(56)+}}\ .
\end{equation}

Eqs.(A.4),(A.5) and (A.6) are the starting--points of calculation in
both, collinear and semicollinear, regions because the full contribution of
(3,4) set of TD can be obtained from (1,2) one by operator $\widehat{O},$
which changes $p_2\leftrightarrow p_-.$

\end{document}